\documentclass[11pt,onecolumn]{article}

\usepackage{lineno,hyperref}
\modulolinenumbers[5]

\usepackage{amsmath}
\usepackage{amssymb}
\usepackage{xspace} \usepackage{ifthen}
\usepackage{array}
\usepackage{xcolor}
\usepackage{url}

\usepackage{graphicx}

\newboolean{figurelist}
\setboolean{figurelist}{true}

\newcommand	{\incfig}	[3]	{\ifthenelse{\boolean{figurelist}}
		{\immediate\write\outstream{fig-#1.pdf}}
	{}
\begin{figure}[!t]
    \includegraphics
				[#2]	{fig-#1}
    \caption{#3}
    \label{fig:#1}
	\end{figure}
	}

\newcommand {\doctable}
	[5] {
	\begin{table}[!t]
\scriptsize
	\caption{#2}
	\label{tbl:#1}
	\centering
	\begin{tabular}{#3}
#4
	\end{tabular}
	\end{table}
	}

\newcommand
	{\comment}
	[1]
	{\textcolor
    	{red}
        {\textbf{#1}}
        }
\renewcommand
	{\comment}
	[1]
	{}

\newenvironment
	{example}
	{
	\begin{description}
	\item [Example:]
	}
	{\end{description}}

\newenvironment{makefigurelist}
	{
\ifthenelse{\boolean{figurelist}}
		{
		\newwrite\outstream
		\immediate\openout\outstream=figure_list
}
		{} }
	{
\ifthenelse{\boolean{figurelist}}
		{
		\immediate\closeout\outstream
		}
		{}
	}

\newcommand{\eqnref}[1] {Eq.\eqref{eq:#1}}
\newcommand{\secref}[1] {\S\ref{sec:#1}}
\newcommand{\tblref}[1] {Tbl.\ref{tbl:#1}}
\newcommand{\figref}[1] {Fig.\ref{fig:#1}}

\newcommand{\ile}[1]{\mbox{$#1$}}
\newcommand{\sss}[2]{#1^{}_{\!_\text{#2}}}
\newcommand{\ssss}[3]{#1^{#3}_{\!_{\text{#2}}}}

\newcommand{\wavelength}{\lambda}
\newcommand{\distanceTarget}{\sss{D}{\text{star}}}
\newcommand{\distanceTargetSquared}{\ssss{D}{\text{star}}{2}}
\newcommand{\rateData}[1]{\mathcal R_{#1}}

\newcommand{\timePulse}{\mathcal T_s}
\newcommand{\BPP}{\text{BPP}}
\newcommand{\PAR}{\text{PAR}}
\newcommand{\numberSlots}{M}
\newcommand{\SBR}{\text{SBR}}
\newcommand{\ratePhotons}[1]{\sss{\Lambda}{#1}}
\newcommand{\efficiencyQuantum}{\eta}
\newcommand{\countPhotonPulse}{K_s}

\newcommand{\bandwidth}{\mathcal W}
\newcommand{\areaAperture}[1]{\sss{A}{#1}}
\newcommand{\numberApertures}{\sss{N}{C}}
\newcommand{\areaCollector}{\areaAperture{C}}
\newcommand{\diameterAperture}[1]{\sss{d}{#1}}
\newcommand{\ang}[1]{\sss{\Theta}{#1}}
\newcommand{\fracBeamOverlap}{\xi}

\newcommand{\powerTx}{\sss{P}{T}}
\newcommand{\dutyCycle}{\delta}

\newcommand{\timeTransmission}{\sss{T}{\text{down}}}
\newcommand{\speedProbe}[1]{\sss{u}{\,#1}}

\newcommand{\powerRec}[1]{\sss{P}{\,#1}}

\newcommand{\dataVolume}{\sss{\mathcal V}{\text{data}}}
\newcommand{\dataVolumeNorm}{\dataVolume/\rateData{0}}
\newcommand{\dataLatency}{\sss{T}{\text{latency}}}

\newcommand{\massRatio}[1]{\sss{\zeta}{\,#1}}
\newcommand{\massRatioE}[2]{\ssss{\zeta}{\,#1}{#2}}
\newcommand{\powerScaleExponent}{k}

\begin{document}

%\begin{frontmatter}
\title{Interstellar flyby scientific data downlink design}
\author{David Messerschmitt \\ Philip Lubin \\ Ian Morrison}
\maketitle

%\title{Interstellar flyby data downlink design\tnoteref{t1}}
%
%
%\author[1]{David G Messerschmitt}
%\address[1]{University of Calfornia at Berkeley,
%Department of Electrical Engineering and Computer Sciences, USA}
%
%\author[2]{Philip Lubin}
%\address[2]{University of California at Santa Barbara,
%Department of Physics, USA}
%
%\author[3]{Ian Morrison}
%\address[3]{Curtin University, International Centre for Radio Astronomy Research, Australia}

%\end{frontmatter}

%\onecolumn

\begin{makefigurelist}

Interstellar distances are vast. Electromagnetic propagation delay grows in proportion to distance, and the propagation power loss grows as the square of distance. These are severe challenges for communication with interstellar spacecraft and probes.

Those who launch such missions may be motivated to achieve scientific returns within a human lifespan or the career of a space scientist or engineer. This leads to the conclusion that such craft or probes must travel at a significant fraction of the speed of light, $c$. This in turn requires great energy resources to impart high kinetic energy, which puts a premium on a spacecraft or probe with a small mass budget.

However, a small total mass implies even less mass allocated to the communication subsystem. This makes it difficult to capture a significant scientific return, which is enabled in part by the volume and reliability of scientific data.

In this tutorial white paper, we discuss the various issues surrounding the design of a communication downlink from a spacecraft or probe at interstellar distances with a constrained mass budget.

\section{Mission architectures}

There are two contrasting types of missions.
The \emph{launch-landing} mission is the more ambitious of the two
and lies beyond our current technological capabilities.
To be viable, a probe needs to approach the speed of light and this invokes relativistic effects.

A more realistic type of mission in the current century is
\emph{launch-flyby} as illustrated in  \figref{flybyMission}, in which a probe
gathers scientific data in the vicinity of a target star
while cruising by at a fixed high speed, and does not require deceleration.
Attention has focused on directed-energy propulsion,
which eliminates the need for a probe to carry fuel for propulsion.
Since the directed-energy acceleration is short-lived relative to the
total mission  duration, the acceleration magnitude needs to be very large, which precludes any possibility of a biological payload.
It is reasonable to assume for purposes of communications that the probe
travels at a constant speed throughout the mission.
It is also generally assumed that the probe speed is on the order of 10\% to 20\% of light speed,
which implies that relativistic effects, while material are also minimal and will be neglected here.
The collection of a finite amount of data during target encounter is followed by downlink transmission
of that data,
so the probe continues to travel at the same constant speed during downlink operation.

\incfig
	{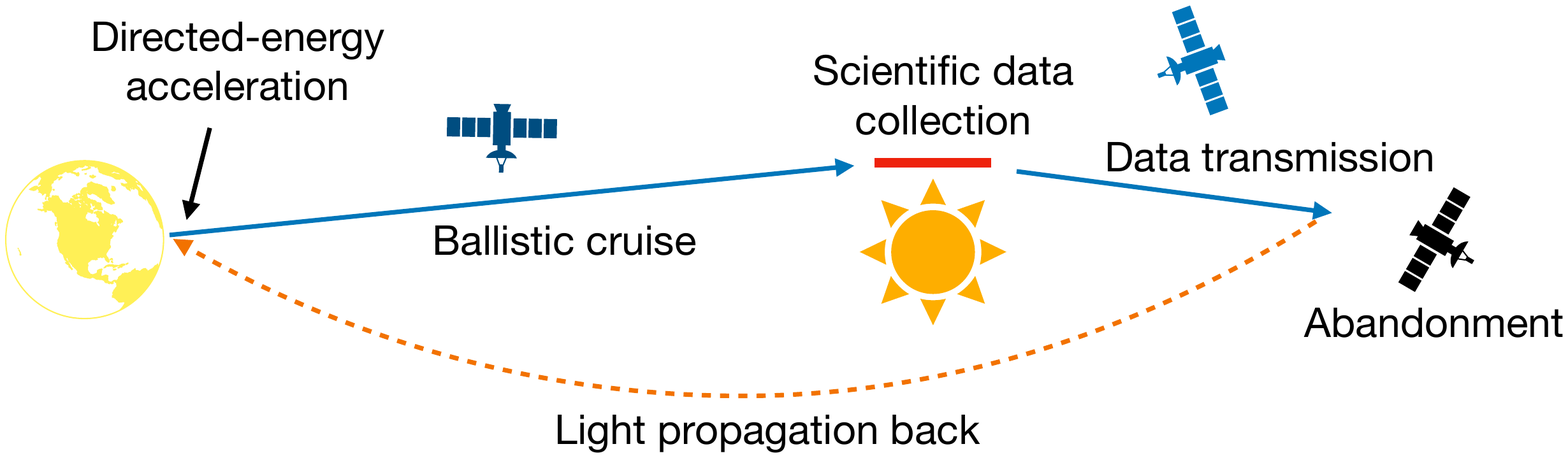}
	{
	trim=20 150 0 250,
    	clip,
    	width=.8\linewidth
	}
	{
	Illustration of the phases of a flyby mission for a probe propelled by
	directed energy from the launch site.
	The goal is to reliably recover
	the collected scientific data at the launch site.
	}

The major cost of a flyby mission is the capital expenditure on a directed-energy beamer at the launch site,
which may be terrestrial, on the moon or, less likely, from a space platform.
Assuming that multiple probes are launched over time,
another significant cost is energy expended during launch.
To reach the high speeds (0.1-0.2 times the light-speed $c$)
that permit travel times to the nearest
stars measured in decades, combined with relatively short acceleration times,
the mass of a probe is expected to be small (perhaps 1 to 1000 grams).
The major communication challenge is realizing a transmitter on the probe
with a small mass budget can reach back from interstellar distances.
It is likely that multiple probes will be conducting this communication concurrently,
and avoiding interference between
the probes is another issue.

Since aggressively constraining the mass budget increases the communication challenge relative to launch-landing missions, flyby with directed-energy propulsion is the application emphasized in this tutorial.
Thus we also specialize the terminology `spacecraft' to `probe', implying a low-mass craft
that follows a ballistic trajectory, that is, unpowered and traveling at constant speed.
Unlike on an uplink,
the small probe mass limits transmit power and there is a premium on reliably returning
as great a volume of scientific data as possible.
Since the most significant challenge is in the design of a downlink, that is emphasized here.
Many of the same issues arise in an uplink to a flyby probe and in communication with
a launch-landing mission.
 
\section{Vast interstellar distances}
\label{sec:distance}

Insight into the impact of distance on the downlink
can be gained from the 
Friis transmission equation \cite{RefNumber871},
which idealizes the propagation power loss as
\begin{equation}
\label{eq:friis1}
\frac{\powerRec{R}}{\powerTx} =  
\frac{1}{{\wavelength^2 \, \distanceTargetSquared} } \cdot
	\areaAperture{T} \areaCollector
	\,,
\end{equation}
where $\powerTx$ is the transmit power and $\powerRec{R}$ is the received power.
The distance to the target  $\distanceTarget$ (where transmission is presumed to occur) is large.
\begin{example}
It is useful to compare this type of interstellar mission with outer planetary exploration of our Solar System.
Pluto averages a distance of \ile{39.5\text{ AU}} from our Sun, 
as compared to \ile{\distanceTarget=4.24\text{ ly}}
to the star \emph{nearest} to our Sun, Proxima Centauri.
The ratio of distances is $6,788$, and this becomes a large power loss to make up when squared.
\end{example}
In \eqnref{friis1} \ile{\{\powerTx,\areaAperture{T},\areaCollector\}} 
represent technological resources we must supply.
Fortunately $\areaCollector$, the size of the collector at the receiver,
is not critical from a mass budget standpoint, whereas the transmit aperture area $\areaAperture{T}$
and transmit power $\powerTx$ do contribute to probe mass.

Fortunately there is another parameter $\wavelength$
which is not a resource and has no direct impact on mass but can be chosen to great advantage.
To a great extent nature's large
$\distanceTarget$ can be compensated by choosing a short $\wavelength$.
The physical reason for this is that apertures of a given area are more directional
at short wavelengths, and thus a short $\wavelength$ can indirectly reduce the required transmit aperture
area $\areaAperture{T}$ and hence probe mass.
\begin{example}
The New Horizons spacecraft, which performed a flyby of planetary objects as
far as Pluto and beyond (the Kuiper Belt), employed a millimeter radio transmission frequency
at \ile{8.438\text{ GHz}} (which corresponds to a wavelength of \ile{\wavelength = 35.53\text{ mm}}).
Suppose we choose for interstellar missions the shortest wavelength that penetrates
our Earth atmosphere efficiently, which is \ile{\wavelength \sim 350{-}400\text{ nm}}.
The ratio \ile{\big(35.53\text{ mm}/400\text{ nm}\big)} is $88{,}825$, which more than compensates for the ratio of
distances.
This is not the full story,
because obtaining a $\powerTx$ and $\areaAperture{T}$ approaching those of
New Horizons is probably not feasable with
a constrained mass budget and near-term technology, 
but it is certainly a great help.
\end{example}
For this and other reasons, this tutorial limits consideration to communication at optical wavelengths.

While \eqnref{friis1} suggests that choosing a large receive collector area $\areaCollector$ can,
like short $\wavelength$, be helpful in overcoming a large $\distanceTarget$,
there are limitations to this based on background radiation.
Fortunately, however, choosing a short $\wavelength$ is directly helpful in this regard also.

\section{Choice of wavelength}
\label{sec:wavelength}

\comment{PL pls check. DM}

The wavelength adopted for communication is a critical choice affects virtually
every other aspect of the system design.
Within the optical, natural sources of background radiation are generally largest at
infrared wavelengths, and fall off dramatically toward the shorter UV wavelengths.
For this reason, UV is strongly preferred for interstellar communication.
An additional advantage of UV is that short wavelengths require commensurately smaller
apertures (see \secref{distance}), 
and the cryogenic cooling of optical detectors is less critical (particularly compared to infrared).

If a terrestrial transmitter or receiver is to be employed, our Earth atmosphere becomes a
severe constraint on wavelength.
In particular the atmosphere is effectively opaque at wavelengths shorter than \ile{350-400\text{ nm}},
which is near the threshold between visible and UV.
In addition, our current optical source and detection technologies are less advanced
in the deep UV because investment-to-date has been driven largely by optical
communication within the atmosphere and within fiber (silica glass is also opaque in the UV).

Generally cosmic sources of background radiation 
(including our Sun, the radiation from
which is scattered within the atmosphere) fall off toward the
lower-wavelength limit of atmospheric transparency, so that is the most favorable region of operation
for terrestrial-based interstellar communication.
This effect is magnified by the characteristic of the target star for exploration by interstellar probes,
since that target is near the probe trajectory and thus becomes a
significant source of interference for communications downlink operation.
In particular, the most favored of targets is Proxima Centauri because it is the nearest star
and also because it is a red dwarf, for which
star-interfering radiation falls off precipitously
at the limits of atmospheric transparency.

For all these reasons we favor the region \ile{\wavelength = 350{-}400\text{ nm}}
for downlink communication from an interstellar probe with a terrestrial receiver.
In this tutorial we adopt the terrestrial receiver assumptions.
With a space-based platform for the receiver, it would be advantageous to move deeper into
the UV, 
where for example \ile{200\text{ nm}} is close to minimum of interfering radiation originating from Proxima Centari.
While our source and detector technologies are less developed at these wavelengths,
this is largely because the primary economic
driver is applications within or through the atmosphere
and not due to fundamental physics.

\section{Layered architecture}
A canonical architecture for achieving energy efficient communications is illustrated
in \figref{layeredArchitecture}.
It is organized into layers, with each layer split between
coordinated functionality across the transmitter and receiver.

\subsection{Functionality}
\label{sec:PHYandTRA}

The partitioning of functionality between the transport layer TRA and 
the physical layer PHY is illustrated in \figref{layeredArchitecture}.
The TRA is logical rather than physical in nature,
and is responsible for composing the structure and content of the transmit signal,
and interpreting the signal detected at the receiver to recover the scientific data.

TRA is divided into two distinct parts.
The modulation coding layer represents
discrete data by a continuous-time intensity waveform at the transmitter,
and interprets the resulting photon events observed in the receiver
in terms of the transmit data (albeit with very poor reliability).
The error-correction coding (ECC) layer introduces a controlled redundancy in the
transmitter and exploits this redundancy in the receiver to recover the
scientific data reliably.

\incfig
	{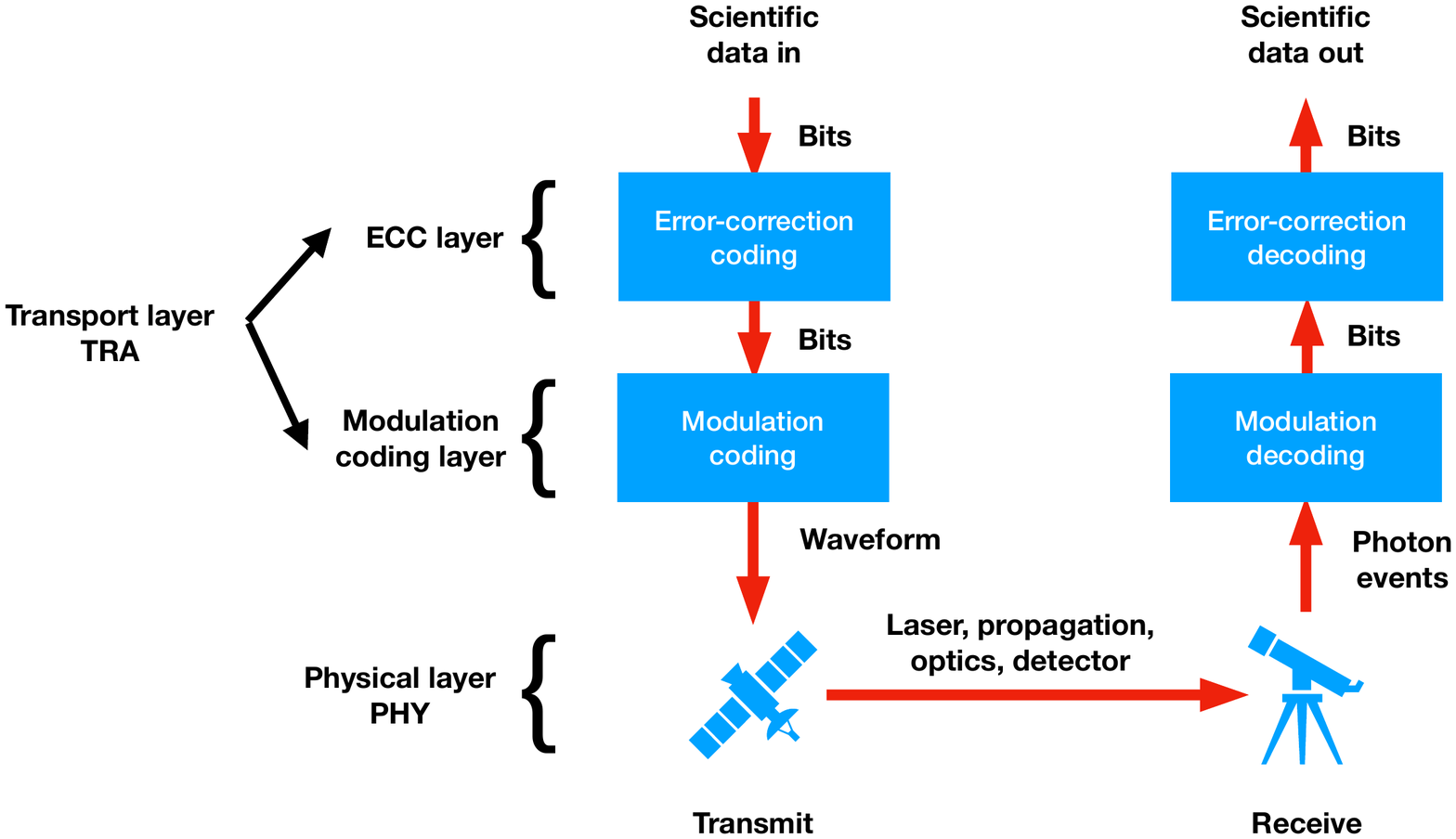}
    {
    trim=0 70 100 70,
    clip,
    width=0.8\linewidth
    }
    {
Coordinated transmit-receive communications architecture.
    Functionality is divided into three layers, with
    each layer divided into a transmit and receive component.
    Logically each layer in the transmitter is coordinated with its counterpart
    in the receiver.
    The physical layer (PHY) includes everything from transmit laser
    to optical detection in the receiver. The transport layer (TRA) (comprised of the modulation coding and ECC layers) provides an intensity-vs-time waveform in the transmitter
    and recovers scientific data from the sequence of photon detection events in the receiver.
    Source: Fig.25 of \cite{RefNumber833}.
    }

\subsection{Goals}
\label{sec:goals}

Two goals for the design are:
\begin{description}
\item{\textbf{Photon efficiency:}}
Accommodate the lowest received signal power (as measured by photon detection rate at the receiver optical detectors), so as to minimize the transmit power, and the transmit and receive aperture sizes.
\item{\textbf{Data reliability:}}
Recover scientific data reliably, as measured by bit error rate.
\end{description}
These two goals are contradictory, as lowering power has a strong tendency to adversely affect reliability,
and if the power is too low then reliability may not be achievable.
The standard approach is thus to achieve the largest lower bound on power over all possible
realizations of TRA,
subject to the constraint that arbitrarily high reliability is achieved.
This theoretical limit is a central result of information theory, and the result is to quantify the lowest power level
for which reliability can theoretically be achieved.
Practical ways of approaching this lower bound within TRA
(and with the cooperation of PHY)
are well developed within information theory and space communications 
and are employed in the following discussion.

\section{Signal photon efficiency}
\label{sec:BPP}

The communications \emph{signal} originates at the probe transmitter,
embeds the scientific information, and is
designed for reliable recovery of that information.
With respect to the signal, there is a clear separation of responsibility between
PHY and TRA:
\begin{itemize}
\item
Let $\ratePhotons{S}$ be the average rate at which photons are \emph{detected} at the output of PHY
in photons per second.
Then PHY should arrange for $\ratePhotons{S}$ to be sufficiently large to extract the scientific
data reliably.
\item
Let $\rateData{}$ be the average rate at which scientific data is extracted in bits per second.
The TRA is primarily responsible for reliable recovery of this scientific data.
Given the large propagation distance and other limitations such as electrical power, it is likely
that \ile{\rateData{} < 1\text{ bits per sec}}.
\end{itemize}

PHY and TRA cooperate to maximize the
\emph{photon efficiency} as measured by the bits per photon ($\BPP$) \cite{RefNumber708,RefNumber656}, 
where\footnote{
In the literature $\BPP$ is sometimes called the \emph{photon information efficiency} (PIE).
}
\begin{equation}
\BPP = \frac{\text{bits reliably recovered per unit time}}{\text{photon detections per unit time}}
= \frac{\rateData{}}{\ratePhotons{S}}
\,.
\end{equation}
Maximizing $\BPP$ is equivalent to minimizing $\ratePhotons{S}$ (and hence signal power)
subject to a constraint on $\rateData{}$.
At the transmitter TRA controls how the power emitted is \emph{modulated} with time.
At the receiver TRA is provided a record of the actual photon detections, essentially
consisting of a sequence of time stamps.
Those time stamps must have sufficient time accuracy as described subsequently.

There is no theoretical constraint on the achievable $\BPP$.
There are, however, practical constraints imposed by limitations in the current technology in use.
Typical values achievable with current technology
are likely to be in the range of
\ile{\BPP \sim 10 \text{ to } 15 \text{ bits/ph}}.
Note that this is multiple bits per detected photon,
and not the other way around.
A successful laboratory demonstration at \ile{\BPP = 13\text{ bits/ph}}
illustrates the possibilities
 \cite{RefNumber834}.

\section{Background radiation}
\label{sec:background}

In this tutorial we make a distinction between \emph{signal},
which is power originating from a communications transmitter,
and \emph{background radiation},
which is spurious sources of photon counts originating from other sources.
We lump these background sources together and call them `radiation', even though
there are non-radiative phenomena among them.
Although background radiation can be overcome to a large extent
by error-control measures (see \secref{reliability}),
in terms of performance measures like scientific data rate and volume,
and in the probe transmitter the
transmit power and aperture size,
limiting and mitigating the background radiation
to the extent possible is always advantageous.

In particular,
the theoretically achievable photon efficiency $\BPP$ is adversely affected by
background unless the signal-to-background power ratio $\SBR$ is sufficiently high
(see \secref{SBR}).
In this tutorial we focus on the communication downlink, since on the uplink it
should be feasible to achieve a high $\SBR$ using a high transmit signal power.
To the extent that
background radiation can be reduced through other measures,
the receiver sensitivity is improved in the sense that lower
signal power becomes acceptable (on either uplink or downlink).

However, even as $\SBR$ becomes high, the inherent quantum
randomness of photon-detection events, especially at optical wavelengths, must be overcome.
This is known as \emph{shot noise}.
When error-control measures are considered later,
an $\SBR$ sufficiently high that background radiation can be neglected is assumed,
but shot noise inevitably remains
(see \secref{reliability}).
Shot noise is perverse in the sense that it increases as the signal power increases,
and thus requires error-control measures to achieve high photon efficiency 
even at high signal powers.

\subsection{Sources of background radiation}
\label{sec:backgroundSources}

The most important sources of background radiation are listed in \tblref{background}.
Many of these are
broadband in nature, and can be largely mitigated
by optical bandpass filtering in the receiver which ideally passes only wavelengths that overlap
the signal bandwidth (see \secref{bandpassFilter}).
Spatially diffuse sources highlight the importance of a receive aperture
that simultaneously gathers signal photons and limits the field
of view (FOV) to
eliminate radiation that is not spatially coincident with the transmitter
(see \secref{choosingNumberApertures}).
It is useful to categorize the types of background radiation in \tblref{background}
into three categories: noise, dark counts and interference.

\doctable
	{background}
    {
    Sources of downlink background radiation}
    {l p{5cm} p{5cm}}
    {
    \hline \\[-2ex]
   & \textbf{Point} & \textbf{Diffuse}
    \\ \hline \\ [-2ex]
    \textbf{Broadband} 
    & 
    Incomplete laser extinguishment
    & 
    Cosmic background
    \\
    && Zodiacal
    \\
    &&Deep star field
    \\[15pt]
    & 
    Target star radiation
    &
    Atmospheric scattering
    \\
    &&(sunlight and moonlight)
    \\[2pt]
    & Dark counts in receive optics and detectors
     \\[20pt]
    \textbf{Bandlimited} 
    & Other transmitters
    & Shot noise
 \\[3pt] \hline
      }

\subsubsection{Noise}

\emph{Noise} is radiation that accompanies, but cannot be separated
from the data signal because of its overlap in
time, wavelength, \emph{and} direction.
Significant sources of noise are the
cosmic background radiation, the deep star field,
zodiacal radiation (due to solar-system dust emissions and scattering),
and sunlight and moonlight scattered in the earth's atmosphere.
The latter varies widely over a \ile{24\text{ hr}}  day
at a terrestrial receiver, and in fact at optical wavelengths
it is impractical to achieve high enough signal power to overcome
sunlight scattering during the daytme \cite{RefNumber833}.
With that conclusion, daytime becomes an outage event 
that significantly reduces the achievable data volume (see \secref{outages}).

\subsubsection{Dark counts}
\label{sec:countsDark}

A very special type of noise is
\emph{dark counts}, which are unwanted photon detections in the
receiver that cannot be attributed to any cosmic source.
If the receive collector is temporarily covered or disconnected,
these dark counts remain.
Generally these dark counts are attributable to black body radiation
from within the receive optics and to spurious photon detection events in
the optical detectors due to thermal effects,
cosmic rays, or other sources.
Later we make a more refined definition of dark counts as
``background which cannot be removed or limited by
optical bandpass filtering'' (see \secref{typesBackground}).
To limit dark counts, it is critical to position optical bandpass filters
late in the optical path, as well as cryogenically chill those filters and everything following,
including the optical single-photon detectors.

\comment{PL pls check. DM}
Following optical bandpass filtering,
the most significant source of background radiation is usually dark counts,
precisely because they cannot be limited by optical bandpass filtering.
Of course this depends on the details of the single-photon optical detection
technology in use,
but technology available today introduces a rate of dark counts that is
quite significant in this application.
The primary reason for this is that a multitude of optical detectors are needed,
and dark counts accumulate across that multitude (see \secref{collector}).
Even an average interval between spurious dark counts of a month or a year,
which renders dark counts insignificant in many other contexts, is a significant 
limitation on the receiver sensitivity in this application to an interstellar downlink.
Technological and conceptual advances can be expected to reduce the prominence of
dark counts, although to an unknown degree.
This is also an area of great opportunity (and necessity) for future research and development.

\subsubsection{Interference}

\emph{Interference}  is unwanted radiation which
is distinguishable from the signal in one
or more physical parameters (time, wavelength, or direction)
and can therefore be partially or
wholly rejected by technological means.
The major interference considered here
is radiation from the target star, which is spatially
separated from probes,
but which may be challenging to reject because of its
close proximity to their trajectory of a flyby mission
or the location of a landing.
Another source is any other artificial transmitters present in or close to
the line of sight, such as other spacecraft or satellites.

For flyby missions there may be multiple probes launched to the same or
nearby targets, and for economic reasons it is likely that a single receive collector
will serve all the probes.
Mutual interference is particularly an issue for directed-energy probes if, after flyby of a common target,
their downlink operations overlap in time (see \secref{downlinkOperationTime}).
In that case, there is a lower limit on the FOV such that all probes are within that view,
and each probe becomes a source of interference to all other probes (see \secref{apertureSize}).
The deliberate combination of such interfering signals in a manner that allows them to be
separated at the receiver, called \emph{multiplexing}, 
is a major issue for some types of multi-probe missions \cite{RefNumber833}.

\subsection{Terrestrial vs space}

A terrestrial location for the receiver has the significant disadvantage of a propagation path
including Earth's atmosphere.
This introduces several challenges that would be avoided in a space-based (satellite or lunar)
platform:
\begin{itemize}
\item
Interference from the target star by choosing an
advantageous wavelength is constrained by any requirement for
atmospheric transparency.
\item
Scattering of sunlight during the daytime and moonlight at night introduce a source of noise.
\item
Signal propagation is susceptible to attenuation by clouds.
\item
Atmospheric turbulence results in signal scintillation and, for larger receive
apertures, likely necessitates the complication of adaptive optics to avoid deep fades.
\end{itemize}
For all these reasons, a space platform for the downlink receiver is advantageous.
However, given the likely necessity for a massive receive collector, this may not be affordable.

\subsection{Signal-to-background (SBR)}
\label{sec:SBR}

The cumulative effect of background is determined by 
the signal-to-background ratio
\ile{\SBR = \ratePhotons{S}/\ratePhotons{B}},
where $\ratePhotons{S}$ and $\ratePhotons{B}$
are respectively the average signal photon rate
(proportional to average input power)
and average background photon rate
at the output of the receiver's optical detector.
Note that $\SBR$ is defined in terms of average power (as opposed to peak power, see \secref{powerPeak}).

To achieve the highest $\BPP$, 
the SBR must be sufficiently high
that background does not
significantly disrupt signal acquisition, synchronization, demodulation and decoding,
and error-correction processes
in the receiver's TRA.
Thus there is a requirement that the $\SBR$ must be above some minimum.
Generally a value \ile{\SBR \ge 4} suffices for that purpose (see Fig.E1 of \cite{RefNumber833}).
The overall background photon rate $\ratePhotons{B}$ limits the receiver sensitivity, specifically
determining the minimum signal photon rate $\ratePhotons{S}$ meeting this minimum 
$\SBR$ requirement.

\subsection{Two types of background}
\label{sec:typesBackground}

There are two components to $\ratePhotons{B}$ that have divergent characteristics,
\begin{equation}
\ratePhotons{B} = \ratePhotons{F} + \ratePhotons{D}
\,.
\end{equation}
The \emph{pre-filtering} background $\ratePhotons{F}$ is introduced in the signal path
(including transmitter, propagation, and receiver) before the optical bandpass filter, and
benefits from bandlimiting.
There are also sources of \emph{post-filtering} background $\ratePhotons{D}$ introduced in the receiver
signal path following optical bandpass filtering.
The most important of these are black body radiation originating in the post-filter optics and
dark counts introduced in optical detectors. 
These two types of background have a duality relationship:
 $\ratePhotons{F}$ but not $\ratePhotons{D}$ benefit from bandpass filtering in the frequency-wavelength domain.
 Meantime $\ratePhotons{D}$ but not $\ratePhotons{F}$ benefit from burst-mode filtering in the time domain.
 This concept is now elaborated.

\subsubsection{Bandwidth control of pre-filtering background}
\label{sec:bandpassFilter}

Optical bandpass filtering in the receiver can eliminate much of the background radiation originating from
broadband cosmic sources and black body radiation introduced in the receiver optics.
Thus in the following we add the following assumption:
\begin{description}
\item{\textbf{Optical bandpass:}}
The optical path in the receiver includes an optical bandpass filter which limits the bandwidth
to the signal bandwidth  $\bandwidth$ (see \secref{bandwidth}).
\end{description}

$\ratePhotons{F}$ can be limited to $\bandwidth$.
Thus, by choice of $\bandwidth$ and coordinated optical bandpass filtering in the receiver,
the size of $\ratePhotons{F}$ can be manipulated to some degree.
These opportunities are limited by the desire for high signal photon efficiency
(which requires bandwidth expansion, see \secref{bandwidth}), and by any technological limitations on
the selectivity of optical bandpass filtering.

\subsubsection{Burst-mode control of post-filtering background}
\label{sec:burst}

$\ratePhotons{D}$ can be limited by an analogous form of filtering in the time domain
called \emph{burst-mode} \cite{RefNumber833},
illustrated in \figref{burstMode}.
During blanking intervals
all background photons can be ignored, thereby reducing the background photon
rate by the \emph{duty-cycle} factor \ile{\dutyCycle = \sss{\mathcal T}{F} / \sss{\mathcal T}{I}}.
This has no net effect on the the pre-filtering background, since even as the
average photon detection rate is reduced by $\dutyCycle$, the signal bandwidth is increased by $\dutyCycle^{-1}$
due to time compression of the bursts, and these two factors precisely offset.

\incfig
	{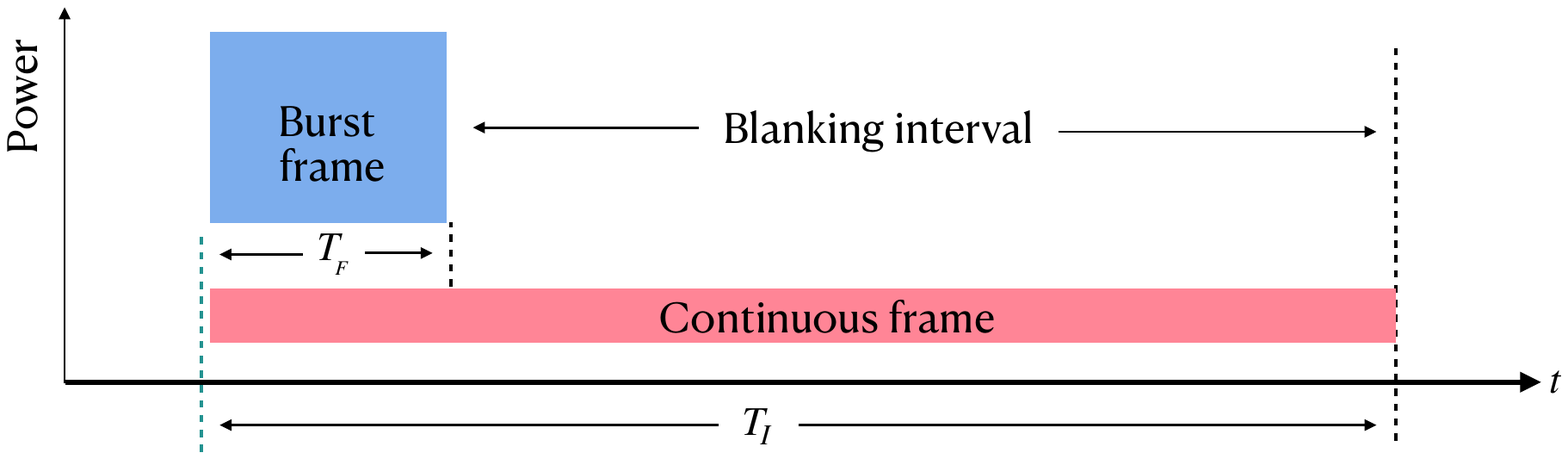}
	{
	trim=0 130 100 300,
    	clip,
    	width=0.9\linewidth
	}
	{
	An illustration of burst-mode transmission, in which a section of transmitted signal of duration
	$\sss{\mathcal T}{I}$ is compressed into a shorter duration 
	\ile{\sss{\mathcal T}{F} < \sss{\mathcal T}{I}}
	at higher power, so that the total energy is not affected.
	This creates a blanking interval known to the receiver,
	during which there can be no signal photon detections and
	any background photons can be ignored.
	}

Overall, burst-mode creates a trade space between the requirement on dark count rate in detectors,
 optical bandpass filtering, and peak power in the transmitter.
The increase in $\bandwidth$ renders the optical bandpass filter less selective, and thus easier to implement.
However the peak transmit power in the transmitter is increased by $\dutyCycle^{-1}$.
This is an indication of the singular importance of peak power in achieving high $\BPP$.
(Another is the high peak power inherent in a modulation coding layers such as 
pulse-position modulation (PPM),
see \secref{PPM}).
In addition, burst mode renders acquisition and synchronization more difficult and more critical
(see \secref{TRArec}).

\section{Transmit/receive aperture size and pointing accuracy}
\label{sec:apertureSize}

A single diffraction-limited aperture is the canonical building block of optical 
electromagnetic transmission and reception.
Either the transmitter or receiver may
be composed from
two or more such apertures (see \secref{trade}),
but for the moment we focus on understanding a single aperture in isolation.
The three parameters that must be coordinated
are illustrated in \figref{pointingBeamCoordination}.
To meet the goal of full target illumination, we must have
\begin{equation}
\label{eq:txApertureConstraint}
\ang{F} > \ang{P} + \ang{T}
\,,
\end{equation}
since otherwise at worst-case pointing angle offset the target may not be fully illuminated
or in the full FOV.
While $\ang{T}$ is determined by the conditions at the target and the uncertainty in its position,
the \ile{\{\ang{F},\ang{P} \}} are design parameters for different probe or receiver subsystems
and must be coordinated to meet this constraint.
The three parameters \ile{\{\ang{F},\ang{P},\ang{T} \}}  are now
discussed  individually.

\incfig
	{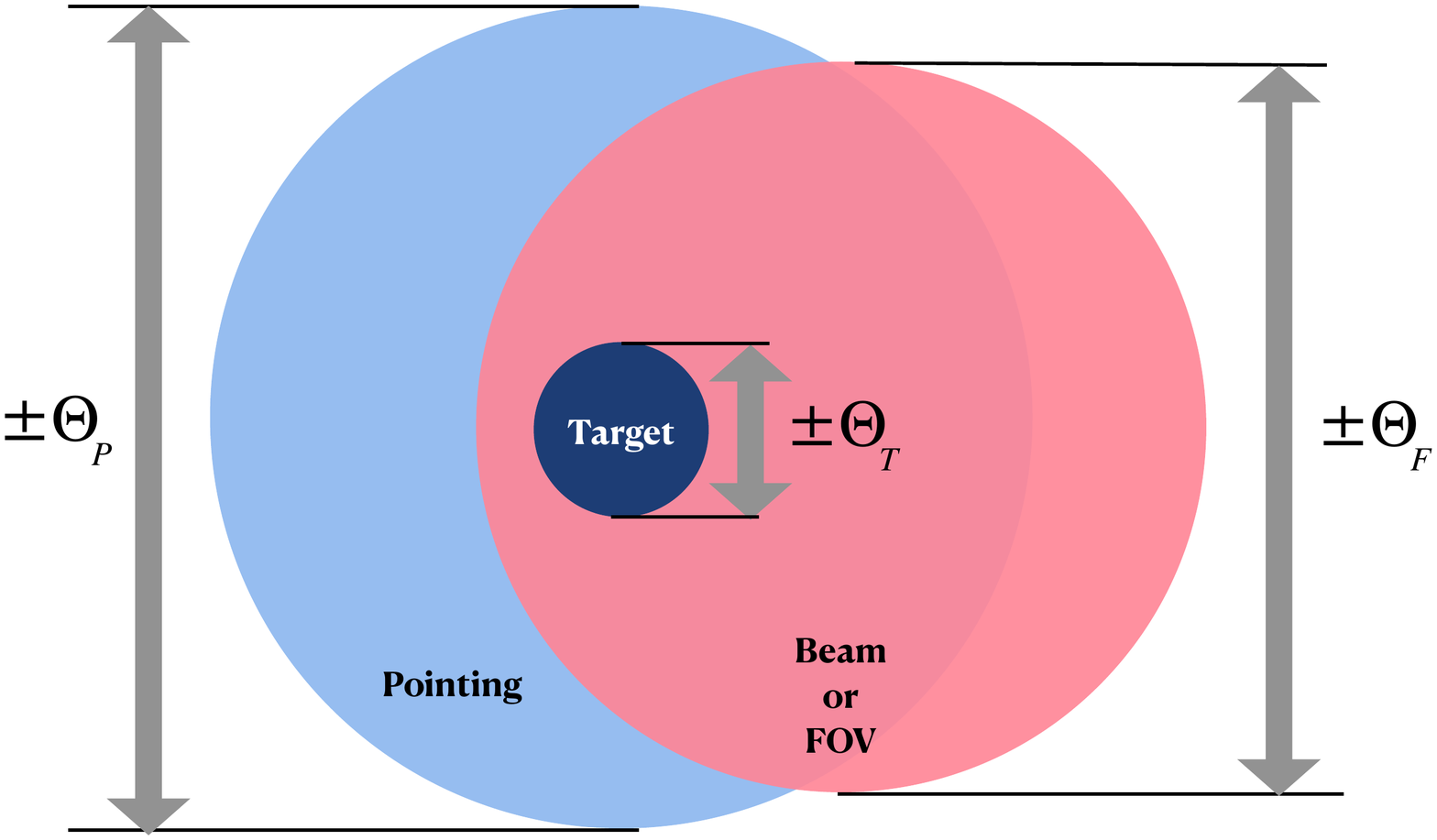}
	{
	trim=0 50 0 50,
    	clip,
    	width=0.7\linewidth
	}
	{
	Illustration of the coordination between the tolerances for pointing accuracy,
	beam (transmit aperture) or field of view (receive aperture), and the target extent.
	These tolerances are defined by their angular variation 
	\ile{\{\pm\ang{P},\pm\ang{F},\pm\ang{T}\}}.
	The goal is to illuminate the entire target at a reasonably uniform flux regardless
	of actual pointing angle within the specified tolerance.
	The circular geometry assumed here can be reformulated for other shapes.
Not accounted for in the case of a receiver is the parallax due to the Earth's orbit, 
	which has a yearly variation and
	can be compensated because it is accurately known.
}

\subsection{Pointing accuracy}

Issues related to pointing accuracy are quite distinctive for space vs terrestrial platforms,
and for transmit beam vs receive FOV.

\subsubsection{Probe transmitter}

In a probe transmitter,
the pointing accuracy tolerance $\pm\ang{P}$ captures how accurately the transmit beam is
aligned with the receiver.
To guide its pointing,
the probe must locate reference stars, one of which is presumably our Sun.

Two forms of alignment are relevant and necessary: The probe attitude
(spatial orientation relative to a standard coordinate system) must be controlled and accurately known,
and second the transmit aperture pointing must be accurately controlled relative to the probe attitude.
In conventional spacecraft these two aspects are likely to be separate (but coordinated) subsystems
due to the need
to align different instruments and propulsion engines in addition to the transmit beam.
In low-mass flyby probes, downlink transmission is likely to be the only function following target encounter,
in which case transmit beam pointing is the only issue.
In this case it may be possible to rely on attitude control
alone, with a rigidly affixed transmit aperture.
In the case where a sail acts as a transmit aperture, accurate attitude control of the sail is necessary.

The achievable pointing accuracy in a low-mass probe is uncertain.
\comment{PL: Pls check the following. DM}
On the one hand it cannot accommodate pointing mechanisms and sensors as bulky as
contemplated in spacecraft with much higher mass,
but on the other hand it avoids vibration originating from various actuators and fluid transfers.
Attitude destabilization results primarily from the momentum transfer of internal electron motion,
any radioactive emissions that may occur from an electrical generator,
emitted photons from attitude control engines and the transmitter, and collisions between
the probe and particles it encounters in the ISM.

\subsubsection{Terrestrial receiver}

In the receiver the pointing accuracy captures
how accurately the FOV is aligned with the target.
In addition to the receive aperture pointing,
a terrestrial receiver is likely to compensate for atmospheric refraction,
and this may require a reference beacon from a space platform.
For a space-platform receiver the issues are qualitatively similar to a probe,
although the mass budget for attitude control and receive aperture pointing would presumably be more generous.

\subsection{Beam divergence or FOV}
\label{sec:apertureFOV}

The size of an aperture determines the angular extent $\pm\ang{F}$ of a beam emitted by the transmitter
or the FOV experienced at the receiver.
Since a larger aperture implies a smaller $\ang{F}$,
\eqnref{txApertureConstraint} places an upper bound on the transmit or receive aperture area.
This limits the signal sensitivity of the communications link.
If the target area can be rendered smaller, or the pointing is rendered more accurate,
the aperture area can be increased and the sensitivity improved.

\subsection{Target uncertainty or size}

A target may be a point, but more likely it has a non-zero angular tolerance $\pm\ang{T}$
attributable to uncertainty as to its location.
For example, astronomical bodies' locations have a level of imprecision,
and likewise a probe trajectory.
Also, the target may have a non-zero extent.
\begin{example}
A receiver may be asked to capture the signals from
multiple probes transmitting concurrently along non-aligned trajectories,
requiring an FOV that covers all those probes.
A target for a probe transmit beam may
consist of our Sun plus Earth orbiting it without knowledge of
 the Earth's location within that orbit,
allowing pointing at the Sun rather than the Earth.
\end{example}

\subsection{Some details and examples}

It is helpful in understanding the challenges faced by a downlink designer
to consider a few numerical examples.

\subsubsection{Beam or FOV radius for a circular axis}

The beam pattern of a diffraction-limited circular aperture has circular symmetry.
A useful measure of its
radius is the angle between the beam axis and the first null in its Fraunhofer diffraction pattern, 
which is to good approximation
\ile{\pm 1.22 \cdot \wavelength/\diameterAperture{}} for wavelength $\wavelength$ and
an aperture diameter $\diameterAperture{}$.
The effective beam spot radius can be set to a fraction \ile{0<\fracBeamOverlap<1} of this first null,
or \ile{\ang{F} = 1.22 \cdot \fracBeamOverlap \wavelength/\diameterAperture{}},
where $\fracBeamOverlap$ is chosen for an acceptable drop in intensity relative to the beam axis.
\begin{example}
Setting \ile{\fracBeamOverlap = 0.3}, the minimum intensity at the limit $\pm \ang{F}$
is 84\% of that at the beam axis.
\end{example}

\subsubsection{Probe transmitter aperture area}

Consider the challenge for a probe transmitter designer.
A larger transmit aperture permits a lower peak and average power, but
the feasible size of the transmit aperture is limited by that aperture's pointing accuracy as well as
by the size of the target defining the possible locations of the receiver.
Pointing accuracy is usually specified in $\mu\text{rad}$ 
($1\ \mu\text{rad}$ is about $0.2\text{ arcsec}$),
where the stated accuracy for planetary missions is less than one $\mu\text{rad}$ at
\ile{\wavelength = 1\ \mu\text{m}} \cite{Lee2001SubmicroradianPS}.
\begin{example}
\comment{PL: Pls check. Apertures are small. DM}
Assume the receive aperture's location is known precisely, or \ile{\ang{T}=0}.
For \ile{\wavelength = 400\text{ nm}}, \ile{\fracBeamOverlap = 0.3},
and pointing accuracy \ile{\ang{P} = \pm 1\ \mu\text{rad}},
we get \ile{\diameterAperture{} < 14.6\text{ cm}}.
Every doubling of $\diameterAperture{}$ is accompanied by a halving of $\ang{P}$.
Also $\diameterAperture{}$ is proportional to $\wavelength$, so a longer $\wavelength$ 
implies a proportionally larger $\diameterAperture{}$.
\end{example}

A larger target (or  \ile{\ang{T} >0}) requires a commensurately smaller aperture.
In practice this can be a significant consideration.

\begin{example}
\comment{PL: Pls check. Apertures are small. DM}
For a probe transmitter,
if the target is considered to be the entirety of the Earth's orbit,
the transmit aperture can be aimed at our Sun without accounting for the orbital position of the Earth.
A target size of \ile{\pm 1\text{ AU}} and a distance of
\ile{4.24\text{ ly}} corresponds to \ile{\ang{T} =  \pm 3.73\ \mu\text{rad}}, which results in
probe transmit aperture
\ile{\diameterAperture{} < 3.1\text{ cm}}.
The aperture size is dominated by target size rather than pointing accuracy.
If it is assumed the probe accurately knows the position of our Sun as well as the
position of the Earth at the time the signal will be received,
this target size can be reduced accordingly with a beneficial increase in transmit aperture size.
\end{example}

\subsubsection{Downlink receiver aperture area}

For the receive aperture, pointing can be rendered more accurate because of the
essentially unlimited mass budget, processing power, etc.
If a receiver were dedicated to a \emph{single} probe, 
the primary issue is the precision with which the probe trajectory is known.
\begin{example}
\comment{PL: Pls check. Apertures are small. DM}
Assume a received aperture pointing accuracy of \ile{\ang{P} = 0.1\ \mu\text{rad}}.
Suppose that any ``scientifically useful'' trajectory
for a star at distance \ile{4.24\text{ ly}} 
has to fall within \ile{10\text{ AU}} of the target star, 
and the position of that star at the point of nearest passage is precisely known.
Then an FOV that covers any useful trajectory becomes 
\ile{\ang{T} = 37.3\ \mu\text{rad}} and the resulting aperture diameter is
\ile{d = 0.4\text{ cm}}.

If, on the other hand, we know the probe trajectory more precisely in terms of
orientation and distance to the target star, the aperture diameter can be increased.
For a passage position relative to the star known within \ile{\pm 0.1\text{ AU}}, the FOV reduces to 
\ile{\ang{T} = 0.37\ \mu\text{rad}} and the resulting aperture diameter is
\ile{d = 31\text{ cm}}.
\end{example}

A typical requirement is to  concurrently receive from multiple probes
with slightly different trajectories.
This increases the required FOV, and reduces the aperture size.
\begin{example}
\comment{PL: Pls check. Apertures are small. DM}
From the receiver FOV perspective, the proper motion of Proxima Centauri
(its changing position relative to the faint background stars) is about
\ile{4\text{ arcsec yr}^{-1}}, which is \ile{19.4\ \mu\text{rad yr}^{-1}}.
If the downlink transmission time is \ile{2\text{ yr}} (see \secref{downlinkOperationTime}), probe trajectories will
track this proper motion over two years, resulting in \ile{\ang{T} \approx 19.4\ \mu\text{rad}}.
This dominates the pointing accuracy considered earlier,
and results in a smaller
receive aperture diameter \ile{\diameterAperture{} < 0.72\text{ cm}}.
However, this is a case where a change in beam shape would be advantageous,
since the star proper motion is linear to good approximation, with a smaller spiral pattern
imposed due to the parallax introduced by the Earth's orbit
(see Fig.8 of \cite{RefNumber833}).
\end{example}

\section{Transmit power and aperture/collector sizes}
\label{sec:collector}

We have seen that often the pointing accuracy and/or target size impose an upper bound on the size
of individual diffraction-limited apertures. 
In fact these apertures
may need to be small if they are to accommodate feasible pointing accuracies for some
typical target sizes.
If additional aperture area is desired, this can be obtained by the incoherent combination of
multiple diffraction-limited apertures.
``Incoherent'' means no attempt to match the phase of the aperture input or output,
but simply put ``powers add''.
This approach has limited appeal on the transmit side because  an attractive alternative
is to incoherently add multiple input signals before applying them to
a single aperture.
\begin{example}
If multiple lasers are required to achieve the desired peak transmit power,
they can be incoherently superimposed optically prior to applying them to a single aperture.
This may have less mass impact than associating an aperture with each laser,
where the superposition occurs in the electrodynamic (propagation) domain.
\end{example}
On the receive side, on the other hand, there is no substitute for a larger total area
to intercept more of the incident signal flux arriving from the transmitter.
This can be achieved by the accumulation of photon detections from multiple apertures.
Thus our development here focuses on the receive side.

\subsection{A collector composed from multiple apertures}
\label{sec:collector}

We make a distinction between an \emph{aperture} and a \emph{collector},
where the collector is composed of multiple apertures as a means of increasing total area
and hence conversion of incident signal flux to a higher detected power.
The simple receive collector architecture shown in
\figref{collectorArchitecture} (as was assumed in \cite{RefNumber833})
composes the collector from multiple diffraction-limited apertures that are combined incoherently.
The individual apertures determine the FOV based on
their aperture area $\areaAperture{R}$, and the number
of apertures $\numberApertures$ determines the total collector area 
\ile{\areaCollector = \numberApertures \areaAperture{R}}.
The incoherent combining, which in the simplest case is obtained by associating an optical
detector with each aperture and accumulating photon
detection events across all apertures, ensures that the FOV is not affected.
Of course a collector so constructed is definitively not diffraction-limited.

\incfig
	{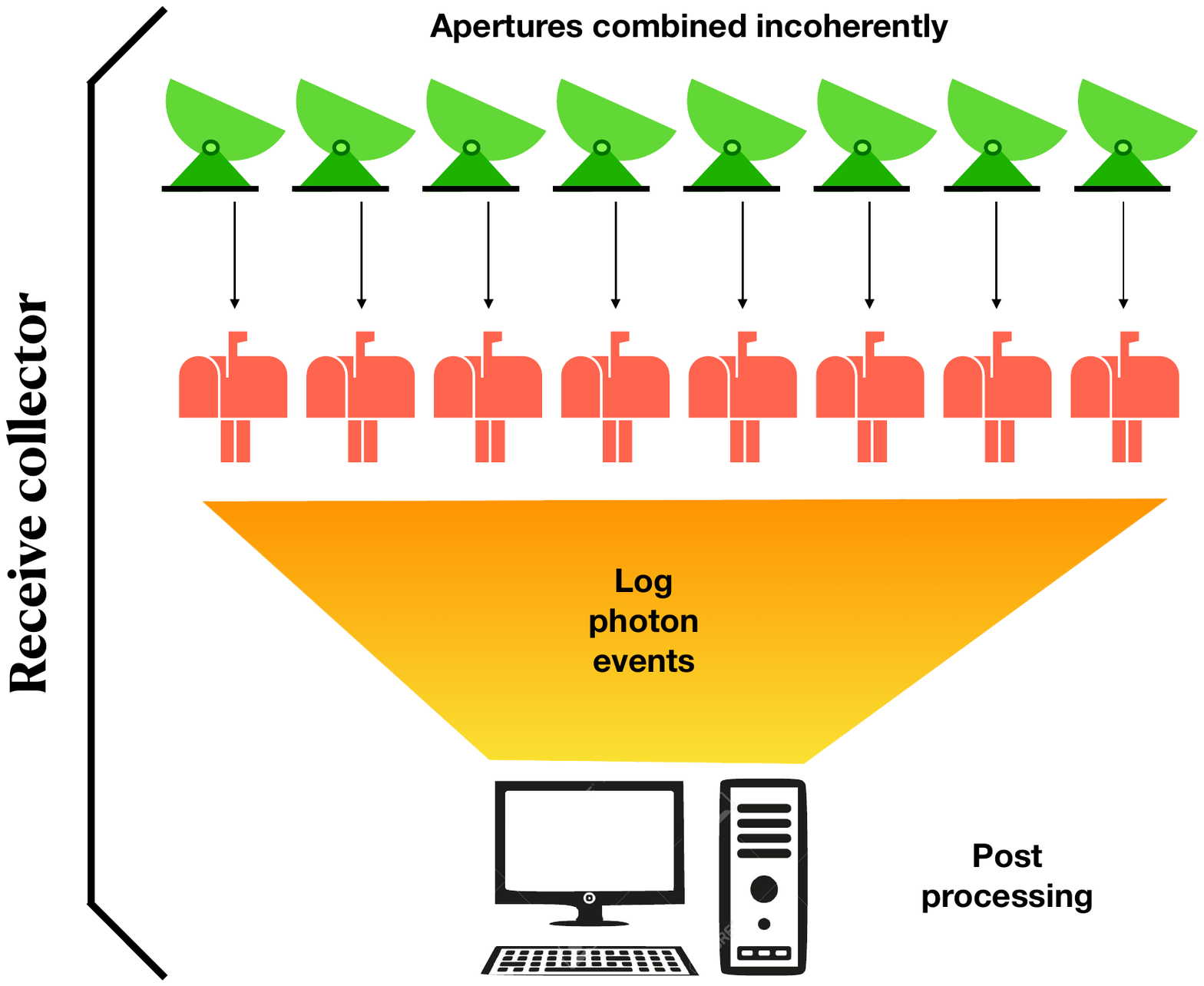}
	{
	trim=50 0 50 0,
    	clip,
    	width=0.8\linewidth
	}
	{
	A receive collector composed of multiple apertures.
	This achieves simultaneously a FOV controlled by the aperture size together
	with a larger total collection area to achieve a 
	signal photon counting rate? $\ratePhotons{S}$ commensurate
	with data rate $\rateData{}$ through the relation \ile{\rateData{} = \ratePhotons{S} \cdot \BPP}.
	For simplicity and without affecting the FOV, the apertures are combined
	incoherently by simply locating a photon-counting optical detector
	(shown as a mailbox icon) at each
	aperture and accumulating their photon counts.
	The resulting collector is not diffraction-limited, but this is fortunately unnecessary at the receiver.
Source: Fig.3 of \cite{RefNumber833}.
}

\subsection{Transmit power and aperture/collector tradeoffs}
\label{sec:trade}

The received power in the Friis transmission equation of \eqnref{friis1} equals the
photon rate times the energy per photon, but an optical detector will have
a quantum efficiency \ile{\efficiencyQuantum<1}, resulting in an average received photon rate
\begin{equation}
\label{eq:friis2}
\ratePhotons{S} =  
\frac
    {\rateData{}}
    {\BPP} =
\efficiencyQuantum \cdot
\frac
	{\areaAperture{T} \areaCollector}
	{h c \, \wavelength \, \distanceTargetSquared} 
\cdot \powerTx
\end{equation}
The factor \ile{\powerTx \areaAperture{T} \areaCollector} creates a trade space
that may significantly impact technological feasibility, probe mass, and cost.
In particular a larger $\areaCollector$ can allow for reduced mass in the probe
through a smaller power-area product \ile{\powerTx \areaAperture{T}}.

\subsubsection{Transmit aperture tradeoffs}

The data rate $\rateData{}$ is proportional to $\wavelength^{-1}$ in \eqnref{friis2}.
Thus as $\wavelength$ is changed and $\powerTx$ is held constant, the
transmit aperture area $\areaAperture{T}$ can be scaled downward as 
\ile{\areaAperture{T} \propto \wavelength}
or the equivalent aperture diameter can be scaled as $\sqrt{\wavelength}$.
This is less favorable than suggested by power considerations alone in \eqnref{friis1}
due to the increasing energy per photon as wavelength decreases.

Short wavelength affects the design and fabrication of both transmit aperture and receive collector,
although the receive collector is less affected as a whole since it need not be diffraction limited.
Depending on area and the chosen technology,
maintaining a transmit aperture close to the diffraction limit at shorter wavelengths will
necessitate tighter fabrication tolerances proportional to the wavelength.
In that event  \ile{\areaAperture{T} \propto \wavelength} does not fully compensate,
and aperture fabrication becomes a greater challenge.

\subsubsection{Receive collector tradeoffs}
\label{sec:collectorTrade}

The aggressive mass budget for a flyby probe suggests a large $\areaCollector$, since this allows for
a commensurately smaller (and lower mass) \ile{\powerTx \areaAperture{T}}.
The following example illustrates that, at interstellar distances and for modest assumptions
on \ile{\{ \powerTx, \areaAperture{T}\}},
$\areaCollector$ can be large (kilometer-scale)
and the required number of apertures composing the collector $\numberApertures$
can also be large (millions to billions).
\begin{example}
For a \ile{\wavelength = 400\text{ nm}},
if we adopt
\ile{\diameterAperture{T} = 14.6\text{ cm}} from a previous example,
then the area of a circular transmit aperture is
\ile{\areaAperture{T} = 167.4\text{ cm}^2}.
Then with \ile{\efficiencyQuantum=0.5}, \ile{\powerTx=10\text{ mW}}, \ile{\distanceTarget=4.24\text{ ly}},
\ile{\rateData{}=1\text{ bits s}^{-1}} and \ile{\BPP = 10\text{ bits ph}^{-1}},
the required receive collector area is \ile{\areaCollector =0.15\text{ km}^2}.
If the receive aperture has diameter \ile{\diameterAperture{R} = 0.72\text{ cm}}
(area \ile{\areaAperture{R} = 0.41\text{ cm}^2})
then the required number of apertures is \ile{\numberApertures=3.75{\times}10^9}.
\end{example}

At the expense of increasing probe mass, the size and scale of the receive collector can be moderated
by increasing the transmit power.
\begin{example}
If the transmit power is increased by $100\times$ to \ile{\powerTx=1\text{ W}}, and
the photon efficiency is increased to $\BPP=15$, and the quantum efficiency is also
increased to \ile{\efficiencyQuantum=0.9}, then
the required collector area decreases to \ile{\areaCollector =566\text{ m}^2}
and the required number of apertures decreases to \ile{\numberApertures=1.38{\times}10^7}.
\end{example}

With a large number of apertures and associated optical detectors, 
the average interval between signal photon detections is large.
This is another indicator of the aggressive requirements on detector dark counts.
\begin{example}
If \ile{\rateData{} = 1\text{ bit s}^{-1}} and \ile{\BPP = 10\text{ bit ph}^{-1}},
the average interval
between photon detections at the collector level is \ile{10\text{ s}}.
A value \ile{\numberApertures=1.38{\times}10^7} corresponds to a \ile{4.4\text{ yr}}
interval at the aperture level.
A typical individual aperture/detector combination detects no signal photons during the entire life of the download operation.
\end{example}

A constraint on this tradeoff is the minimum $\SBR$ requirement at the optical detector (see \secref{SBR}).
The trade space is therefore influenced by the impact of 
\ile{\{\areaAperture{R},\numberApertures,\areaCollector\}}
on background, which is now considered.
Our ability to reduce \ile{\powerTx \areaAperture{T}}
is constrained by the need for a sufficiently large
signal at the receiver to overcome the background radiation and hence achieve a sufficiently large $\SBR$.

\subsubsection{Background vs receive aperture and collector size}
\label{sec:choosingNumberApertures}

Pre-filtering background of cosmic origin is usually unresolved, meaning that the
source of background has a fixed areal density within the FOV of the aperture.
For such background, and for a single receive aperture with area $\areaAperture{R}$,
the background power reaching an optical detector is
not dependent on $\areaAperture{R}$ (see App.B of \cite{RefNumber833}).
An explanation for this (arguably counter-intuitive) statement is that as $\areaAperture{R}$ increases, the total background appearing as
incident flux to the aperture decreases due to the decreasing FOV, but the area of that flux
captured by the aperture increases, and these two factors offset.

There may also be resolved point sources of background, such as radiation from a nearby star
that falls within the FOV (the target star will usually fall in this category),
for which the incident flux does not depend on $\areaAperture{R}$ and the total captured background
thus increases 
in proportion to $\areaAperture{R}$. 

Much like unresolved sources of cosmic background,
post-filtering background is also independent of $\areaAperture{R}$, 
but for the very different reason that it is introduced following the aperture.

The canonical collector architecture of \figref{collectorArchitecture} does not affect $\SBR$.
To see this,
let $\sss{\SBR}{R}$ and $\sss{\SBR}{C} $ be the $\SBR$ at the aperture and collector level.
Then since both signal photon rate $\ratePhotons{S}$ and 
background photon rate $\ratePhotons{B}$ (both at the level of a single aperture and its
dedicated optical detector) accumulate incoherently,
\begin{equation}
\label{eq:SBRcollector}
\sss{\SBR}{C} = \frac{\numberApertures \ratePhotons{S}}{\numberApertures \ratePhotons{B}} = 
\frac{ \ratePhotons{S}}{ \ratePhotons{B}} = \sss{\SBR}{R}
\,.
\end{equation}
Thus the $\SBR$ constraint can be set on the basis of a single receive aperture.
With the notable exception of point sources of radiation, $\SBR$ is independent of
$\areaAperture{R}$, depending instead primarily on the size of the post-filtering background,
which is determined primarily by the technology of the optical detector and its temperature.

Receive collector design can thus be summarized as:
\begin{itemize}
\item
Choose $\areaAperture{R}$ on the basis of the desired FOV, which places an upper bound
(see \secref{apertureFOV}).
\item
Manipulate the signal level at the aperture level through the choice of \ile{\powerTx \areaAperture{T}}.
Knowing the background level, this signal level can be chosen to meet the minimum acceptable $\SBR$,
thus beneficially minimizing \ile{\powerTx \areaAperture{T}}.
Choosing the largest acceptable $\areaAperture{R}$ consistent with FOV is beneficial in 
allowing the smallest possible \ile{\powerTx \areaAperture{T}}.
\item
Choose $\numberApertures$ so that the signal photon rate
\ile{\numberApertures \ratePhotons{S}} at the level of the entire collector 
suffices to support the
data rate $\rateData{}$ for the expected photon efficiency $\BPP$,
namely \ile{\numberApertures \ratePhotons{S} \cdot \BPP = \rateData{}}.
\end{itemize}

\subsubsection{Receive collector improvements}

It is likely that the canonical collector of \figref{collectorArchitecture} can be improved on
through invention and technology development.
A couple of approaches have been suggested \cite{RefNumber833},
but surely there are others.
\comment{PL: Pls check. DM}
One is to arrange for a larger $\areaAperture{R}$, which allows a commensurate decrease in
\ile{\powerTx \areaAperture{T}} while maintaining fixed signal level.
\begin{example}
If $\areaAperture{R}$ is limited by the larger FOV required for coverage of multiple probes,
replicating the receiver multiple times, each dedicated to one probe, would allow
$\areaAperture{R}$ to be increased.
This would be expensive, but it may be possible to accomplish the same effect by
using a larger $\areaAperture{R}$ in conjunction with imaging (as in an optical telescope).
In that case different probes could be associated with different pixels, each with an effectively
larger aperture.
\end{example}
Another approach is to achieve \ile{\sss{\SBR}{C} > \sss{\SBR}{R}} in \eqnref{SBRcollector},
which also allows \ile{\powerTx \areaAperture{T}} to be decreased.
\begin{example}
An architecture in which there are fewer than $\numberApertures$ optical detectors will
reduce the post-filtering background in the denominator of \eqnref{SBRcollector}.
This might be accomplished by sharing a single optical detector among as many apertures as possible.
\end{example}

\section{Outages}
\label{sec:outages}

Operation during daylight is most certainly ruled out, and at night weather
events may create a cloud cover that blocks the signal.
These events are called \emph{outages}.
Weather-related outages are a significant disadvantage to using optical wavelengths,
and can be avoided by employing a relay satellite to recover the
scientific data and retransmit it at radio wavelengths to a terrestrial radio receiver.
This approach has been used in planetary optical communications \cite{Lee2001SubmicroradianPS},
but is less attractive for interstellar optical communications because of the massive receive collector 
that would be required in space due to much greater signal propagation distance.

For a terrestrial-based receive collector, outages are a major consideration in
achieving reliable recovery of scientific data.
Accommodating outages is a well-known challenge in wireless communication, and there are standard
mitigation techniques.
The design of the ECC has to take outages into account, and the achievable reliable data rate $\rateData{}$
will be adversely affected (see \secref{outageCoding}).

\section{Downlink operation time}
\label{sec:downlinkOperationTime}

For a flyby mission, the total duration of the downlink operation is
expected to be on the order of years.
There are several compelling reasons for this assumption, as now summarized.
This relates to the design of the mission, including the mass and speed of the probe
and the duration of the operation of the downlink data transmission $\timeTransmission$.
The following is a summary of results, which are covered more thoroughly in \cite{messer2022mass}.

\subsection{Data volume and latency tradeoff}
\label{sec:volumeLatency}

While this tutorial emphasizes metrics of performance like data rate and photon efficiency,
these are internal communications design issues of little concern to domain scientists
who are waiting for the data return from the probe mission
in order to enable their scientific interpretation of that data.
These science stakeholders are primarily interested in three issues:
The data return reliable (addressed in \secref{reliability}), the \emph{data volume} $\dataVolume$
(total number of data bits returned), and the \emph{data latency} $\dataLatency$
(elapsed time from probe launch to return of the data in its entirety).
 In other words, how much data do we get back reliably, and how long do we have to wait for that data?

Two mission design parameters for a flyby mission with directed-energy propulsion are
the  duration of downlink transmission $\timeTransmission$ and the probe mass ratio
$\massRatio{P}$,
which is the ratio of the actual probe mass to some baseline value.
The significance of $\massRatio{P}$ is that if the launch beam and power remain fixed
across multiple probe launches, the speed of the probe is directly affected by the probe mass,
and hence by  $\massRatio{P}$.
In particular the speed decreases as $\massRatioE{P}{-1/4}$ with increasing mass,
and the total launch energy increases as $\massRatioE{P}{3/4}$.

Although the mission design parameters \ile{\{\massRatio{P},\timeTransmission\}} can be varied to 
manipulate the mission performance metrics \ile{\{\dataVolume,\dataLatency\}},
they should not be chosen arbitrarily.
Rather, they should be
jointly optimized to achieve the most favorable \ile{\{\dataVolume,\dataLatency\}}
tradeoff.
The impact of \ile{\{\massRatio{P},\timeTransmission\}} on \ile{\{\dataVolume,\dataLatency\}} is slightly complicated,
but can be summarized as:
\begin{itemize}
\item
The nature of directed energy propulsion is that if the launch beam remains fixed
(except possibly for the time duration of probe acceleration),
a larger $\massRatio{P}$ results in a
lower cruise speed for the probe $\speedProbe{P}$.
\item
 This increases the travel time to the target, and this increases
 $\dataLatency$.
\item
This results in a smaller accumulation of distance-squared propagation delay,
and thus allows $\rateData{}$ to decrease more slowly during downlink transmission, and
this in turn increases $\dataVolume$.
\item
For fixed $\timeTransmission$, the probe has traveled less distance from the target
during downlink operation, the maximum propagation delay back to the launch site
is smaller, and this reduces $\dataLatency$.
\item
A larger probe mass budget for communications
allows its electrical power and hence transmit power to be increased,
and also possibly the transmit aperture to be increased in area.
These benefits increase $\rateData{}$ and thereby $\dataVolume$.
\end{itemize}
While the travel-time increase is deleterious, all the other impacts are beneficial.
Trading these off leads to an optimum point.

\subsubsection{Optimal volume-latency tradeoff}
\label{sec:volumeLatencyTradeoff}

The design of a data link, which conveys data reliably from probe to launch site,
involves a number of interacting considerations such as wavelength,
transmit aperture, receive collector, background radiation,
modulation, and coding.
For purposes of mission design,
all these considerations can be wrapped into a single parameter, 
a baseline data rate $\rateData{0}$ at the beginning of
downlink operation assuming \ile{\massRatio{P}=1}.
The total data volume \ile{\dataVolume \propto \rateData{0}}, and thus
it is convenient to use the normalized volume \ile{\dataVolumeNorm}
(which is dimensioned in time) as a performance metric to guide the
choice of \ile{\{\massRatio{P},\timeTransmission\}}.

 The tradeoff between $\dataVolumeNorm$ and $\dataLatency$
  is plotted in \figref{efficientFrontierVsFixedMassRatio} \cite{RefNumber833,messer2022mass}.
 There exists a feasible region of operation \ile{\{\dataVolumeNorm,\dataLatency\}}, which
 is shaded\footnote{
This version of the efficient frontier assumes \ile{\rateData{0} \propto {\massRatio{P}}^{\!\!\!2}}.
 The shape depends on the particulars of this scaling, which is an important characteristic
 of any probe technology \cite{messer2022mass}.
}
 in \figref{efficientFrontierVsFixedMassRatio}.
 Points on the lower boundary of
 this region, called the \emph{efficient frontier}\footnote{
 This terminology is borrowed from a similar concept in financial portfolio theory \cite{merton1972analytic}.
 It is a special case of the \emph{Pareto frontier} (Pareto optimization is widely employed in
 various engineering disciplines \cite{jakob2014pareto}).
 },
 constitute all the advantageous mission operating points.
That boundary yields the smallest possible $\dataLatency$ for a given $\dataVolumeNorm$,
or the largest possible $\dataVolumeNorm$ for a given $\dataLatency$.
This assumes that 
\ile{\rateData{0} \propto {\massRatio{P}}^\powerScaleExponent}
for \ile{\powerScaleExponent=2}
(alternative values for $\powerScaleExponent$ are considered in \cite{messer2022mass}).
Numerical values along the efficient frontier are listed in
\tblref{efficientFrontierExp2}
along with the associated optimal set of parameters.
\begin{example}
If we place a limit on the latency of \ile{36.5\text{ yr}}, then the maximum normalized data volume
is $\dataVolumeNorm=10^9\text{ s}$.
At an initial data rate of \ile{\rateData{0} = 0.1\text{ bits s}^{-1}},
the maximum data volume is 
\ile{\dataVolume = 100\text{ Mb}} (megabits) or  \ile{12.5\text{ MB}} (megabytes).
\end{example}

\doctable
	{efficientFrontierExp2}
	{
	Efficient frontier and related parameters for a scaling exponent \ile{\powerScaleExponent = 2. }
	}
	{ccccccc}
	{
	 \\ \hline \\[-2ex]
 $\log_{10} \big[ \dataVolumeNorm \big]$ & $\dataLatency\text{ in yr}$ 
 & $\massRatio{P}$ & $\speedProbe{P}/c$ & $\timeTransmission\text{ in yr}$
 \\[-2ex]
    \\ \hline
\\$0.$ & $8.5$ & $0.0016$ & $1.$ & $0.0124$
\\$1.$ & $8.74$ & $0.0016$ & $1.$ & $0.128$
\\$2.$ & $9.89$ & $0.0033$ & $0.834$ & $0.308$
\\$3.$ & $11.5$ & $0.0088$ & $0.653$ & $0.437$
\\$4.$ & $13.5$ & $0.0236$ & $0.51$ & $0.611$
\\$5.$ & $16.1$ & $0.0636$ & $0.398$ & $0.846$
\\$6.$ & $19.4$ & $0.172$ & $0.31$ & $1.16$
\\$7.$ & $23.7$ & $0.469$ & $0.242$ & $1.57$
\\$8.$ & $29.3$ & $1.28$ & $0.188$ & $2.11$
\\$9.$ & $36.5$ & $3.52$ & $0.146$ & $2.81$
\\$10.$ & $45.8$ & $9.67$ & $0.113$ & $3.73$
\\$11.$ & $57.8$ & $26.7$ & $0.088$ & $4.92$
\\$12.$ & $73.2$ & $73.6$ & $0.0683$ & $6.46$
\\$13.$ & $93.2$ & $204.$ & $0.0529$ & $8.45$
	}

\subsubsection{Choosing a point on the efficient frontier}

Choice of a mission operation point somewhere on the efficient frontier provides flexibility
in setting mission priorities.
There are several compelling reasons to consciously select different
operating points along the efficient frontier for different missions 
sharing a common launch infrastructure:
\begin{itemize}
\item
 Mission designers can consciously
 prioritize large $\dataVolumeNorm$ or small $\dataLatency$.
 \item
 Different probes may carry different types of instrumentation,
 and these impose different mass and data volume requirements.
 \item
 There will likely be an evolution of probe technology over time.
 Early probes may emphasize technology validation
 with low $\dataLatency$ (and hence small $\dataVolumeNorm$),
 while later probes may emphasize scientific return with larger $\dataVolumeNorm$
 (and hence larger  $\dataLatency$).
 \item
 There may be missions to different targets at different distances
(within the solar system and interstellar), significantly changing the
 possible range of $\dataLatency$.
 \end{itemize}
 Generally, mission designers will seek to maximize an objective function that
 combines volume and latency objectives.
 Only points along the efficient frontier need be considered in any such optimization.
 
 Also illustrated in \figref{efficientFrontierVsFixedMassRatio} as the dashed curve is
 a set of possible mission operation points when
a baseline value \ile{\massRatio{P}=1}  is chosen and only $\timeTransmission$ is varied.
This approaches the minimum-latency horizontal asymptote as \ile{\timeTransmission \to 0}
and a maximum-volume vertical asymptote as \ile{\timeTransmission \to \infty}.
This arbitrary choice of  \ile{\massRatio{P}} permits operation at exactly 
 one point on the efficient frontier through a
 judicious choice of $\timeTransmission$;
 otherwise, it is inferior.
More generally, achieving an arbitrary operating point on the efficient frontier requires a
coordinated choice of \ile{\{\massRatio{P},\timeTransmission\}} rather than constraining
$\massRatio{P}$ in this manner.
The efficient frontier is the lower envelope of all such curves varied over all values of $\massRatio{P}$.

\incfig
	{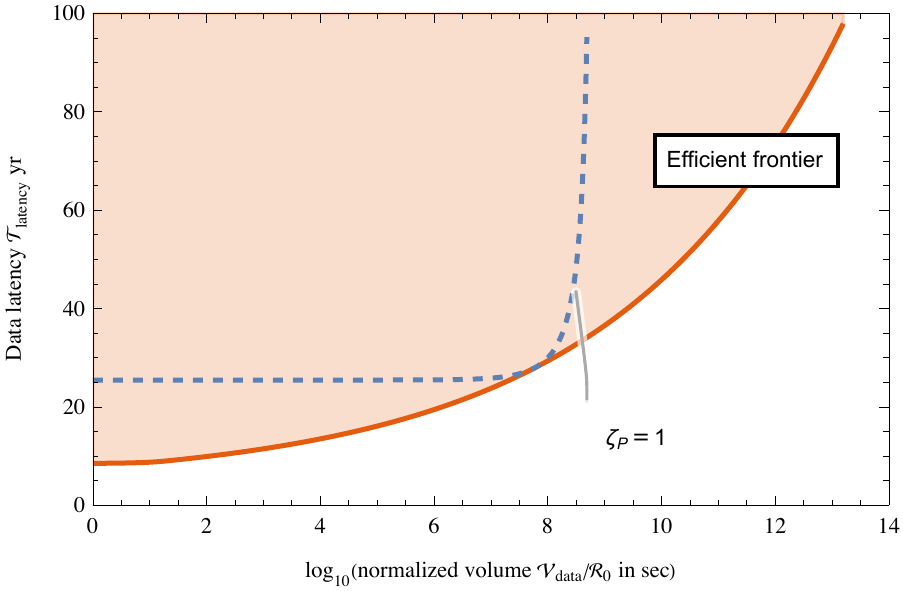}
	{
	trim=0 0 0 0,
    	clip,
    	width=0.7\linewidth
	}
	{
Plots of data latency $\dataLatency$ (in years) vs the log of the normalized data volume
	$\dataVolumeNorm$ (in seconds) where 
	$\rateData{0}$ is the data rate (in bits per second) 
	at the beginning of downlink transmission
	(data rate declines from there as the square of propagation distance) for mass ratio \ile{\massRatio{P}=1}.
	$\dataVolume$ (in bits) is found by multiplying by the assumed value for $\rateData{0}$.
	Any volume-latency mission operating point within the shaded region is feasible.
	The lower boundary of this region, called the efficient frontier, is an efficient operating point in the 
	sense of maximizing the volume for a given latency,
	or minimizing latency for a given volume.
	The set of operation points obtained by fixing \ile{\massRatio{P} = 1}
	and varying downlink operation duration $\timeTransmission$ are shown as a dashed curve.
}
 
\subsubsection{What changes along the efficient frontier}
How does the mission design change along the efficient frontier?
This can be summarized as, from lower to higher normalized data volume $\dataVolumeNorm$:
\begin{itemize}
\item
The mass ratio $\massRatio{P}$ increases, the probe mass increases, and the probe speed decreases.
\item
The data latency $\dataLatency$ increases, although this penalty is the most favorable possible.
\item
The downlink operation duration $\timeTransmission$ increases.
A useful rule of thumb is that it approximately equals 9\% of the probe transit time from launch to encounter
with the target star.\footnote{
This value depends on assumptions about how probe power and aperture area scale with $\massRatio{P}$.
See \cite{messer2022mass} for further details.
}
For example, with a transit time of \ile{30\text{ yr}}, the downlink would operate for \ile{2.7\text{ yr}}.
\item
The launch energy increases, since the probe is accelerated for a longer period of time.
\item
The total distance the probe
travels to the end of downlink operation does not vary significantly along the efficient frontier; 
that is, it always travels a total distance
\ile{\approx 1.09\cdot \distanceTarget} to the point it suspends downlink operation.
\end{itemize}
In summary,
along the efficient frontier the mission design is dominated by the distance-squared law of
propagation loss, and thus the total distance traveled to the suspension of downlink operation does not vary much.
Higher $\dataVolume$ can be achieved by reducing the probe speed, giving the probe a longer
period to transmit its data before the distance-square loss becomes dominant.
There is a direct mission cost penalty for achieving larger $\dataVolume$ in increased launch energy.

\subsection{Other considerations in downlink duration}

There are considerations beyond optimizing the volume-latency tradeoff that
strongly suggest a downlink operation time on the order of years following target-star encounter 
and scientific data collection:
\begin{itemize}
\item
Electrical generation
based on a reservoir of `fuel' or a battery is likely inconsistent with mass objectives.
Rather, it is likely to be based on a radiative-thermal source or the
(necessarily slow) extraction of energy from the interstellar medium, which are
low power but persistent.
At the very least the electrical generation capability has to exceed the multi-decade transit time
to the target star.
\item
For reliable data recovery in the face of random outages, the downlink operation time 
should be at least $50\times$ the longest outage time (see \secref{outageCoding}).
Depending on the location of a terrestrial receiver, 
outages due to weather events may last a significant number of days,
implying a downlink operation on the order of years.
\end{itemize}

\section{Modulation code: PPM}
\label{sec:PPM}

The modulation coding layer in \figref{layeredArchitecture} 
converts bits supplied by the ECC layer into a light intensity waveform to be generated
by the transmit laser.
A modulation coding layer commonly in use is \emph{pulse-position modulation} (PPM), 
which is illustrated in \figref{PPM}.
PPM combined with an appropriate ECC layer can achieve close to theoretical constraints
on photon efficiency, subject to assumptions about peak and average power.

\incfig
	{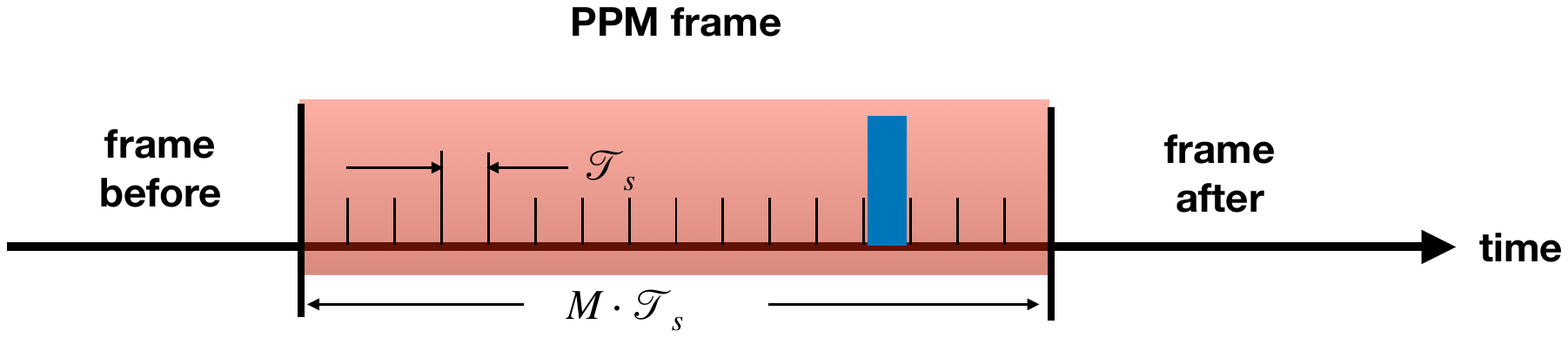}
	{
	trim=50 230 50 250,
    	clip,
    	width=1.0\linewidth
	}
	{
	Pulse-position modulation divides the received signal power at the optical detector into
	frames, where each frame is further subdivided into $\numberSlots$ slots each of duration $\timePulse$.
	The convention is that
	the received signal power is all within a single slot within the frame, which communicates
	$\log_2 \numberSlots$ bits of ``raw'' information (if the slots are equally likely to be so-energized).
	The average number of photons detected for each energized slot is $\countPhotonPulse$,
	and this is also the average number of photons detected for the whole frame.
	The peak-to-average power ratio is thus \ile{\PAR = \numberSlots}.
	}

\subsection{Theoretical photon efficiency of PPM}
\label{sec:PPMcapacity}

Information theory provides a least upper bound
(over all possible error-correction coding (ECC) schemes, see \secref{ECC})
on the achievable photon efficiency of PPM \cite{RefNumber798},
\begin{equation}
\label{eq:PPMcapacity}
 \BPP \le \alpha \big[ \countPhotonPulse \big] \cdot \log_2 \numberSlots
\text{     where     }
\alpha \big[\countPhotonPulse \big] = \frac{1 - \text e^{-\countPhotonPulse} }{\countPhotonPulse}
\,.
 \end{equation}
 A Poisson distribution governing the count of detected photons is assumed, and the
 parameter $\countPhotonPulse$ equals the average number of detected photons
 in an energized PPM slot
(which is one out of $\numberSlots$ total slots in each PPM frame,
 $(\numberSlots-1)$ of which are not energized).
Since \ile{0 < \alpha \big[\countPhotonPulse \big] < 1} monotonically approaches 
\ile{\alpha \big[\countPhotonPulse \big] \to 1}
as \ile{\countPhotonPulse \to 0},
this suggests that $\countPhotonPulse$ should be small to achieve the highest $\BPP$.

A similar least upper bound over all possible transmitter-receiver combinations that
utilize photon-counting detection (and 
therefore doesn't
presuppose PPM) corresponds to \ile{\alpha[\countPhotonPulse] = 1}.
Thus PPM is photon efficient, but only if we can arrange for \ile{\alpha[\countPhotonPulse] \sim 1},
or equivalently a small value for $\countPhotonPulse$.

For any value of $\BPP$, the PPM parameters \ile{\{\countPhotonPulse,\numberSlots\}}
can be coordinated to achieve that $\BPP$ efficiently  (see \secref{optimization}).
The resulting 
\ile{\{\countPhotonPulse,\numberSlots\}} are plotted in \figref{photonStarvation}
and the resulting values of $\BPP$ are tabulated in
\tblref{parametersVsBPP2} for representative values of $\numberSlots$
chosen to be powers of two (a typical implementation convenience).
Also shown is the resulting value of \ile{\alpha[\countPhotonPulse]}, which quantifies the shortfall
attributable to specializing the photon-counting scheme to PPM.
Two additional columns are included, and are discussed later.

\incfig
	{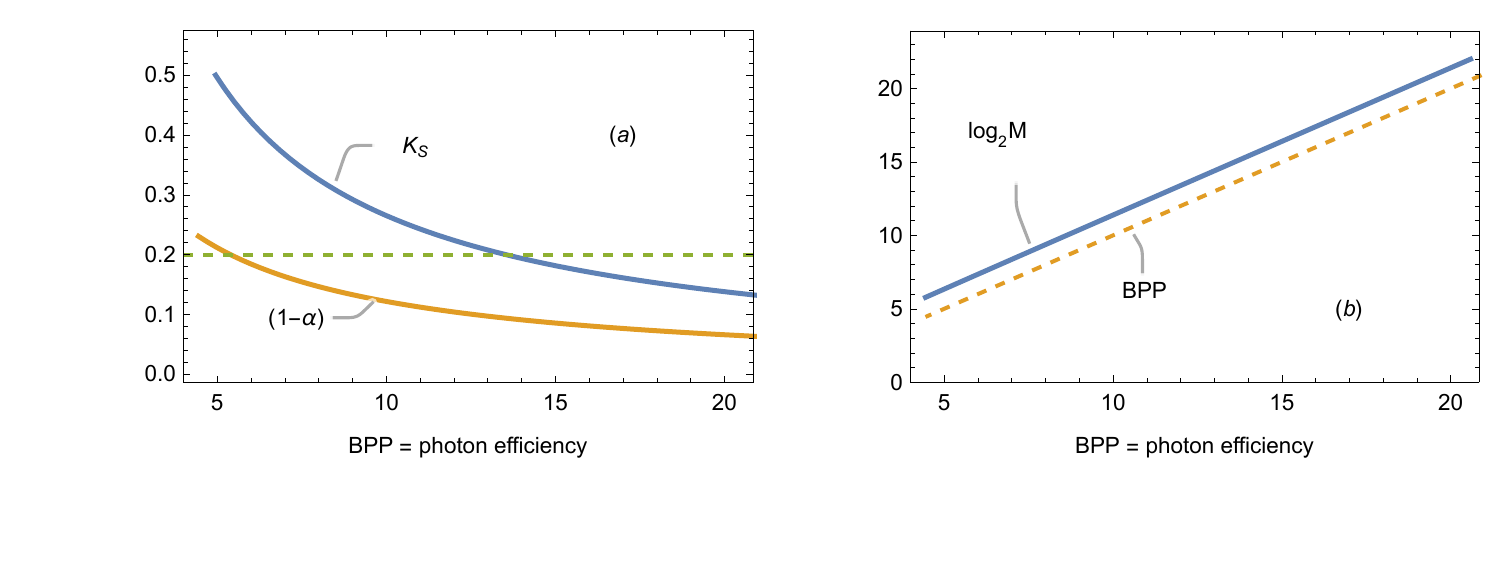}
	{
	trim=30 30 0 0,
    	clip,
    	width=1\linewidth
	}
	{
At the theoretical least upper bound $\BPP$ on PPM photon efficiency,
	PPM parameters \ile{\{\countPhotonPulse,\numberSlots\}}
	that achieve that bound efficiently are plotted.
	Distinct dimensionless values are plotted, with the curves rather than the axes labeled.
	(a) The value of $\countPhotonPulse$ illustrates photon starvation.
	The value \ile{\countPhotonPulse = 0.2}, which is a reasonable choice in the region of interest, is
	plotted as a dashed line.
	Also plotted is the decrease in photon efficiency \ile{\big(1-\alpha[\countPhotonPulse]\big)}
	relative to the best possible transmitter-receiver utilizing
	photon counting detection but without limiting to PPM, which is on the order of 10\% to 20\%.
	(b) The number of bits per PPM frame $\log_2 \numberSlots$ is compared to $\BPP$
	in the dashed line.
That \ile{\BPP < \log_2 \numberSlots} by a small margin is another 
	indication of PPM's modest inefficiency relative to a general photon-counting scheme.
}

\doctable
	{parametersVsBPP2}
	{
	Optimal choice of PPM parameters for \ile{\numberSlots = 2^m}
	}
	{ccccccc}
	{
	 \\ \hline \\[-2ex]
 $\BPP$ & $\countPhotonPulse$ & $m$ & \ile{\numberSlots = 2^m} &
 $\alpha[\countPhotonPulse]$ & $\timePulse \rateData{}$ & $\bandwidth/\rateData{}$
    \\ \hline \\ [-2ex]
$0.815$ & $2.17$ & $2$ & $4.$ & $0.408$ & $0.443$ & $2.26$\\
$1.75$ & $1.19$ & $3$ & $8.$ & $0.584$ & $0.261$ & $3.83$\\
$2.71$ & $0.837$ & $4$ & $16.$ & $0.678$ & $0.142$ & $7.06$\\
$3.68$ & $0.646$ & $5$ & $32.$ & $0.737$ & $0.0744$ & $13.4$\\
$4.66$ & $0.527$ & $6$ & $64.$ & $0.777$ & $0.0384$ & $26.$\\
$5.65$ & $0.445$ & $7$ & $128.$ & $0.807$ & $0.0196$ & $50.9$\\
$6.64$ & $0.385$ & $8$ & $256.$ & $0.83$ & $0.00999$ & $100.$\\
$7.63$ & $0.34$ & $9$ & $512.$ & $0.848$ & $0.00506$ & $197.$\\
$8.62$ & $0.304$ & $10$ & $1020.$ & $0.862$ & $0.00256$ & $391.$\\
$9.62$ & $0.275$ & $11$ & $2050.$ & $0.874$ & $0.00129$ & $775.$\\
$10.6$ & $0.251$ & $12$ & $4100.$ & $0.884$ & $6.5{\times}10^{-4}$ & $1540.$\\
$11.6$ & $0.231$ & $13$ & $8190.$ & $0.893$ & $3.27{\times}10^{-4}$ & $3060.$\\
$12.6$ & $0.214$ & $14$ & $1.64{\times}10^{4}$ & $0.9$ & $1.64{\times}10^{-4}$ & $6080.$\\
$13.6$ & $0.199$ & $15$ & $3.28{\times}10^{4}$ & $0.907$ & $8.26{\times}10^{-5}$ & $1.21{\times}10^{4}$\\
$14.6$ & $0.186$ & $16$ & $6.55{\times}10^{4}$ & $0.912$ & $4.15{\times}10^{-5}$ & $2.41{\times}10^{4}$\\
$15.6$ & $0.175$ & $17$ & $1.31{\times}10^{5}$ & $0.917$ & $2.08{\times}10^{-5}$ & $4.81{\times}10^{4}$\\
$16.6$ & $0.165$ & $18$ & $2.62{\times}10^{5}$ & $0.922$ & $1.04{\times}10^{-5}$ & $9.58{\times}10^{4}$\\
$17.6$ & $0.156$ & $19$ & $5.24{\times}10^{5}$ & $0.926$ & $5.23{\times}10^{-6}$ & $1.91{\times}10^{5}$\\
$18.6$ & $0.148$ & $20$ & $1.05{\times}10^{6}$ & $0.93$ & $2.62{\times}10^{-6}$ & $3.81{\times}10^{5}$\\
$19.6$ & $0.141$ & $21$ & $2.1{\times}10^{6}$ & $0.933$ & $1.31{\times}10^{-6}$ & $7.61{\times}10^{5}$
	}

While the optimum value of $\countPhotonPulse$ depends on the $\BPP$ we aspire to achieve,
in the range of interest available with near-term technology
the value \ile{\countPhotonPulse \sim 0.2} is a good approximation.
That there is a fraction of one detected photon per pulse on average,
which is called \emph{photon starvation}, may be surprising.
The implication of this is that most frames actually experience zero photon detections (see \secref{TRArec}).
Achieving the $\BPP$ that we aspire to requires a large value for $\numberSlots$.
\begin{example}
If we seek to achieve the least upper bound of \ile{\BPP \le 15\text{ bits/ph}},
this requires \ile{\numberSlots = 2^{17} = 131,072} in \tblref{parametersVsBPP2}.
In this case 17 bits have to be passed from TRA to PHY to specify which PPM slot to energize within each PPM frame.
This large $\numberSlots$ renders  synchronization to be challenging.
The resulting \ile{\alpha[0.175]= 0.917}, 
and thus the photon efficiency bound is 8.3\% below a general photon-counting
modulation scheme with the same peak-to-average ratio.
\end{example}
For concrete choices and implementations of error-correction codes,
the best choice of $K_s$ will generally be slightly higher and
the achieved efficiency will generally be lower than this theoretical bound (see \secref{reliability}).

\subsection{Peak power}
\label{sec:powerPeak}

For any direct detection (photon-counting) receiver the
photon efficiency is bounded by \ile{\BPP \le \log_2 \PAR},
where $\PAR$ is the peak-power to average-power ratio within a frame \cite{RefNumber799}.
This is consistent with PPM, where \ile{\PAR = M} as given in \eqnref{PPMcapacity}.
Thus we must have \ile{\PAR > 2^\BPP}, which increases exponentially with $\BPP$.
Further burst mode control of post-filtering background increases
$\PAR$ by another factor of $\delta^{-1}$.
Practical limits on $\PAR$ for semiconductor lasers are likely to severely limit
available $\BPP$ and/or our ability to control post-background radiation.
This motivates an interest in optical detectors with very low dark count rates (see \secref{countsDark})
and pulse-compression technologies (see \secref{dispersion}).

\subsection{TRA-to-PHY interface in transmitter}

At the transmitter, for each PPM frame the TRA specifies the appropriate timeslot, which requires
\ile{\log_2 \numberSlots} bits.
Typically $\numberSlots$ is chosen to be a power of two, so that this is an integral number of bits.
Then observing that the total number of photon detections per frame
(as well as per slot) is $\countPhotonPulse$, and from \eqnref{PPMcapacity},
\begin{equation}
\label{eq:fracSciBits}
\frac{\text{ average scientific data bits per PPM frame }}{\text{TRA to PHY bits per PPM frame}} =
\frac{\countPhotonPulse \cdot \BPP}{\log_2 \numberSlots} \le
\countPhotonPulse \alpha[\countPhotonPulse]
\,.
\end{equation}
Thus each PPM frame represents considerably fewer scientific data bits than
the number of bits actually passed from the ECC to the modulation code for that same frame
in \figref{layeredArchitecture}.
The remainder of those bits constitute redundancy, which is used for error control (see \secref{ECC}).
\begin{example}
For the parameter choices in the last example,
the value of \eqnref{fracSciBits} is 0.166.
Thus only 16.6\% of the bits passed from TRA to PHY represent scientific data, 
and 83.4\% of the bits represent redundancy for error control.
\end{example}
It is important to recognize that each PPM frame does not stand on its own,
but rather the added redundancy is processed across large blocks of PPM frames
so that statistical averaging can be exploited (see \secref{ECC}).

\subsection{PHY-to-TRA interface in receiver}
\label{sec:TRArec}

At the receiver, TRA is provided a list of all photon detection events by PHY.
This is a list of time stamps identifying the time of the detection
and any other available information such as detected photon wavelengths
or pulse shapes at the detector 
(which may be relevant to vetoing some spurious counts of non-cosmic origin).
This list constitutes the primary scientific record of the mission, 
which may conceivably be processed multiple times in the future as algorithms are improved.

Due to the inherent unreliability of the PHY, because it is subject to quantum effects in the signal 
as well as spurious photon counts due to background radiation,
the output of the PHY at the receiver is only stochastically related to the input to the PHY at the transmitter.
TRA in the receiver starts by performing \emph{synchronization}, 
which estimates the PPM frame and slot boundaries, and is in itself a
difficult step that may require offline processing over a significant portion of the downlink duration.
Synchronization is rendered more critical as well as more difficult by both PPM and
by burst mode imposed on top of PPM (see \secref{burst}).

Assume that synchronization has been successfully acquired.
Then for each received PPM frame, TRA may observe one of three conditions:
\begin{enumerate}
\item
There are one or more photon detections, and all such photon detections fall in the same PPM slot.
\item
There are two or more photon detections, but they do not all fall in the same PPM slot.
This has to be the result of spurious photon count(s) in non-energized slot(s).
\item
There are zero photon detections within the entire PPM frame.
This is typically due to the inherent stochastic nature of signal-photon detection.
\end{enumerate}
The second and third cases are called an \emph{erasure}, since in either case it is
impossible to associate a unique PPM time slot with the reception.
In the first case, the photon detections may be in the correct time slot (as determined by transmission),
or a different time slot (which is called an \emph{error}).
Erasures due to photon starvation are commonplace.
\begin{example}
Continuing the past examples,
and assuming no background radiation,
if \ile{\countPhotonPulse = 0.175} then the probability of an erasure (applying the Poisson distribution) is
\ile{\text e^{-0.175} = 0.839}.
Thus one or more photons are detected on average for only 16\% of frames, and on average the remaining
83.9\% of frames are erasures.
\end{example}
The additional information inherent in erasures (as opposed to errors) can be exploited in the subsequent
decoding to improve the reliability.
Reliable recovery of scientific data  in spite of erasures and errors is achievable due to the inherent redundancy
deliberately introduced at the transmitter, reflected in a number of bits per frame passed from TRA to PHY
that exceeds the average number of scientific data bits decoded from each frame.

Based on the sequence of unambiguously detected time slots per PPM frame, combined with (mostly) erasures, 
TRA can proceed with the recovery of the scientific data
through a process that is typically complex and computationally burdensome.
Along with synchronization,
this process is likely to occur offline and in non-real time.

\subsection{The canonical on-off pulse}

The basic building block for PPM that is of primary concern to PHY is defined at the receiver optical detector,
rather than at the transmitter. 
The received power at the detector should be confined to one PPM time slot of duration $\timePulse$.
In that case, for $(\numberSlots-1)$ out of the $\numberSlots$ slots in one PPM frame there is no signal energy, and thus
any photon detections in those slots are attributable to background radiation.
It is the total energy of the pulse that determines the average number of photon detections, 
and not the detail of how power varies within the time slot.

The procedure for choosing the number of apertures $\numberApertures$ in a collector
was described in \secref{choosingNumberApertures} in terms of average photon rate.
When specializing to PPM, this procedure can be equivalently performed with respect to
a single canonical pulse peak power,
with the goal of making $\numberApertures$ sufficiently large that 
 an average photon detection count $\countPhotonPulse$ per pulse
is realized at the collector output.

\subsection{Time slot duration}
\label{sec:countPerPulse}

With knowledge of the PPM parameters \ile{\{\countPhotonPulse,\numberSlots\}} and the resulting $\BPP$,
the time slot duration $\timePulse$ is readily inferred.
Equating two representations of the volume of scientific data associated with a single PPM frame,
\begin{equation}
\label{eq:slotTime}
\timePulse \numberSlots \cdot \rateData{} = \countPhotonPulse \cdot \BPP
\,,
\end{equation}
we see that, for fixed $\numberSlots$, \ile{\timePulse \propto {\rateData{}}^{-1}}.
\begin{example}
Using the same parameters as in previous examples, and assuming \ile{\rateData{} = 1\text{ bits s}^{-1}},
we find from \eqnref{slotTime} that the required \ile{\timePulse = 31.4\ \mu\text{s}}.
Increasing $\rateData{}$ to $10\text{ bits s}^{-1}$ reduces the slot duration to $3.14\ \mu\text{s}$.
\end{example}

Another implication of \eqnref{slotTime} follows from \eqnref{PPMcapacity},
\begin{equation}
\label{eq:slotTimeData}
\timePulse \rateData{} = \countPhotonPulse \alpha[\countPhotonPulse]  \cdot
 \frac{\log_2 \numberSlots}{\numberSlots}
 \,.
 \end{equation}
This value, which is tabulated in \tblref{parametersVsBPP2}, is useful for determining $\timePulse$ for any $\rateData{}$.
Since $\countPhotonPulse$ doesn't vary too much as $\BPP$ is changed (see \figref{photonStarvation}),
as $\BPP$ is increased the
value of $\timePulse \rateData{}$ (and hence $\timePulse$ for fixed $\rateData{}$) 
decreases roughly in proportion to 
\ile{\big( \log_2 \numberSlots \big)/\numberSlots}.
However the total frame time \ile{\numberSlots \timePulse}
increases roughly as \ile{\log_2 \numberSlots}, and the number of scientific data bits
\ile{\numberSlots \timePulse \rateData{}}
represented by each frame increases the same way.
Thus \eqnref{fracSciBits} and \eqnref{slotTimeData} are consistent.
\begin{example}
Using the same parameters as in previous examples, and assuming \ile{\rateData{} = 1\text{ bits s}^{-1}},
we find from \eqnref{slotTime} that the time slot decreases from   \ile{\timePulse = 992\ \mu\text{s}} 
to \ile{31.4\ \mu\text{s}} as the photon efficiency is increased from \ile{\BPP = 10\text{ bits/ph}} to
\ile{15\text{ bits/ph}}.
The number of slots per PPM frame increases from \ile{\numberSlots = 2,673} to $86,573$.
\end{example}

\subsection{Burst mode and PPM}

\newcommand{\timeFrameInterval}{\sss{\mathcal T}{F}}

Another consideration affecting the choice of PPM parameterization
is management of post-filter background through burst-mode
(see \secref{burst}).
Suppose the time interval between PPM frames is $\timeFrameInterval$,
which relates directly to the scientific data rate $\rateData{}$.
Normally PPM is thought of as full duty-cycle \ile{\dutyCycle = 1}, for which
\ile{\timeFrameInterval = \numberSlots \timePulse}.
However, if instead \ile{\numberSlots \timePulse \ll \timeFrameInterval} is deliberately chosen,
that leaves blanking intervals between PPM frames,
and the average post-filtering background photon detection count
\ile{\ratePhotons{D} \numberSlots \timePulse} during each PPM frame (see \secref{burst})
is reduced.
That in turn permits a smaller pulse energy while maintaining an adequate $\SBR$ (see \secref{SBR}),
thereby reducing the transmit power burden on the transmitter.
Of course the number of apertures $\numberApertures$ in the receive collector 
has to be commensurately larger (see \secref{collectorTrade}),
but overall burst-mode beneficially permits a shift in resources from transmitter to receiver.

There is of course a price to pay.
Assuming that the pulse energy is fixed (and hence the average transmit power is also fixed),
there are two ways to reduce the PPM frame time $\numberSlots \timePulse$,
and each has adverse consequences:
\begin{itemize}
\item
Reducing $\numberSlots$ will reduce the photon efficiency $\BPP$ and hence $\rateData{}$.
\item
Reducing $\timePulse$ will increase the peak power.
\end{itemize}
Overall burst-mode creates a trade space among peak transmit power, data rate,
and the partitioning of resources between transmitter and receiver.
As the dependencies are multi-faceted, these tradeoffs are best addressed in
the context of a numerical model that permits parameter exploration.

\subsection{Optical bandwidth}
\label{sec:bandwidth}

Since the signal is a linear superposition of canonical pulses, the
optical bandwidth of the signal $\bandwidth$ is the same as that of a single
pulse.
In particular it is related to $\timePulse$
through the uncertainty relationship \ile{\bandwidth \sim \timePulse^{-1}}.
The metric
\begin{equation}
\label{eq:expansion}
\text{bandwidth expansion} =  \frac{\bandwidth}{\rateData{}} = \frac{1}{\timePulse \rateData{}}
\end{equation}
tells us the factor by which optical bandwidth is expanded relative to the data rate.
Although it appears to depend on $\timePulse \rateData{}$,
this is misleading as seen in \eqnref{slotTimeData}.
In fact it depends only on the PPM parameters, and
its value is also tabulated in \tblref{parametersVsBPP2}.
\begin{example}
For the parameters of earlier examples the bandwidth expansion factor is 31,816.
Thus, the optical bandwidth is \ile{\bandwidth = 31.8\text{ kHz}}  for \ile{\rateData{} = 1\text{ bits s}^{-1}}
and the bandwidth expansion is $3.18 \cdot 10^4$.
\end{example}
If the scientific data rate was increased or burst-mode PPM was employed (see \secref{burst}), 
this bandwidth would increase accordingly.
Larger optical bandwidth beneficially reduces the required optical bandpass filter selectivity,
but also adversely affects the background radiation admitted to the optical detector (see \secref{background}).

\subsubsection{Theoretical limits on bandwidth expansion}

A general conclusion across all of communications is that if the
objective is to achieve higher energy efficiency, this requires an increase in
bandwidth expansion.
$\bandwidth/\rateData{}$ is plotted in \figref{opticalLimitsRho} for PPM with an optimum choice of
parameters based on \eqnref{expansion}.
Also shown is the quantum limit, which removes the assumption of a photon-counting receiver
(and hence PPM) and allows the transmitter and receiver to manipulate quantum states directly
(see \secref{holevo}).
For a given $\bandwidth/\rateData{}$, higher BPP can be achieved at the quantum limit,
indicating the possibilities for more advanced future technologies that manipulate quantum states
rather than simply counting photons.
In interstellar communciation, the penalty of larger $\bandwidth$ is a larger pre-filtering background radiation,
which then requires a larger signal level to achieve an acceptable $\SBR$ (see \secref{SBR}).
If post-filtering background is dominant, this is less relevant.

\incfig
	{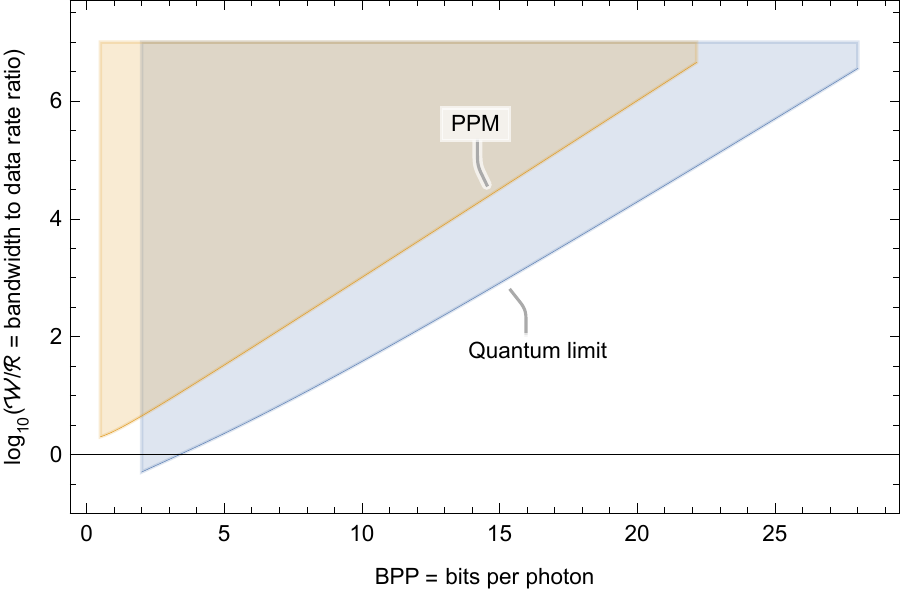}
	{
	trim=0 0 0 0,
    	clip,
    	width=.7\linewidth
	}
	{
For communication at optical wavelengths, a log plot of the theoretical minimum
	bandwidth expansion \ile{ \bandwidth/\rateData{}} for a given photon efficiency $\BPP$.
	Operation within the shaded areas is feasible, and
	achieving greater efficiency requires a larger bandwidth.
	Shown is the quantum limit derived in \secref{holevo}, and the less favorable
	tradeoff for a PPM modulation code with optimum choice of parameters based on photon counting detection.
}

\subsection{Purposeful time dispersion}
\label{sec:dispersion}

A significant issue is the peak power required to support a high $\BPP$ and large $\numberSlots$,
because the peak power required to maintain fixed $\countPhotonPulse$
grows as $\timePulse$ decreases (see \secref{countPerPulse}).
This peak power can be beneficially decreased without affecting total energy
through the deliberate introduction of \emph{time dispersion}.

Suppose two pulses both have constant power and the same bandwidth $\bandwidth$ but different time durations, 
\begin{equation}
\big| p_1 (t) \big|^2 = P\,,\ 0 < t < T \text{   and   } \big| p_2 (t) \big|^2 = P \cdot \bandwidth T \,,\ 0 < t < 1/\bandwidth \,,
\end{equation}
where $p_1 (t)$ has a longer time duration than is necessary for that bandwidth (\ile{\bandwidth T \gg 1}).
Then the two pulses have equivalent energy, but the minimum
duration pulse $p_2 (t)$ has greater power by the factor $\bandwidth T$.
Thus $p_1 (t)$ can be transmitted at lower power, and since the two pulses have the
same bandwidth a subsequent linear filtering de-dispersion operation over that bandwidth can
produce at the detector a higher power pulse of shorter duration and equal energy.
\begin{example}
If the time slot duration is \ile{\timePulse = 1\ \mu\text{s}}, then the required optical bandwidth is
\ile{\bandwidth = 1\text{ MHz}}.
A deliberately dispersed pulse with time duration \ile{1\text{ ms}} and the same bandwidth can be generated by
sweeping the wavelength over \ile{1\text{ MHz}} during the \ile{1\text{ ms}} duration.
This pulse is called a \emph{chirp}.
After de-dispersion the pulse width at the optical detector is \ile{1\ \mu\text{s}} with the peak power
increased by a factor of $10^3$.
\end{example}
The de-dispersion step may be located at any convenient point in the signal path, including the transmitter
or the receiver.
This technique is used in optics to generate short pulses with very high power \cite{RefNumber1006,RefNumber1005},
and may similarly be beneficial here.
However, the generally wider pulses required at the low bit rates that may be feasible
in interstellar communication
presents an obstacle for these techniques \cite{RefNumber833}.

\section{Achieving reliability}
\label{sec:reliability}

It is an understatement to say that PHY in isolation is unreliable, given that on the order of
80\% of PPM frames are erased (contain zero detected photons).
This seeming `flaw' is necessary for achieving a high $\BPP$, but it must be overcome within the TRA.
The justification for this partitioning of responsibility falls back on the inherent conflict between high $\BPP$
and the requirement for reliable recovery of data.
The design is thus partitioned into two phases:
\begin{description}
\item{\textbf{PHY:}}
High energy efficiency is obtained by embedding information in the \emph{timing} of narrow pulses,
since adjusting the timing of canonical pulses within a PPM frame incurs no energy penalty.
High energy efficiency is also achieved through photon starvation, 
with the side effect of poor reliability within PHY.
\item{\textbf{TRA:}}
Reliability is obtained by adding redundancy
in the transmitter TRA, and utilizing that redundancy
(the structure of which is known in detail by the receiver) 
in the receiver TRA to overcome the high erasure rate in the PHY.
\end{description}

\subsection{Error-correction coding}
\label{sec:ECC}

We saw earlier that redundancy makes up the bulk of the data that is passed from TRA to PHY in the
transmitter.
The redundancy insertion and exploitation cannot take place on a
frame-by-frame basis, as the reliability would still be poor.
Rather, the stochastic nature of the quantum channel
(with or without background radiation)
has to be countered by averaging over long time intervals, exploiting the law of large numbers.

To make use of that law, blocks of $n$ symbols are defined as vectors
in the transmitter TRA and the receiver TRA.
PHY then associates the $i$th output vector with the $i$th input vector,
\begin{align*}
\overrightarrow{\mathbf X_i} \to \overrightarrow{\mathbf Y_i}
 \,,\text{ where }
&\overrightarrow{\mathbf X_i} = \big[ X_{ i \cdot n+1}, X_{i \cdot n +2}, \dots X_{(i+1) \cdot n} \big]
\\
&\overrightarrow{\mathbf Y_i} = \big[ Y_{ i \cdot n+1}, Y_{i \cdot n +2}, \dots Y_{(i+1) \cdot n} \big]
\,.
\end{align*}
A block of scientific data is mapped into a transmit block $\overrightarrow{\mathbf X_i} $.
The number of feasible values for $\overrightarrow{\mathbf X_i} $ is far greater than
the cardinality of the input scientific data represented within that block,
where the difference is the added redundancy.
At the receiver the distances between the stochastic reception $\overrightarrow{\mathbf Y_i} $
and all possible ``legitimate'' values of  $\overrightarrow{\mathbf X_i} $ are determined,
and the scientific data corresponding to the closest such distance is chosen, in a process known as \ile{decoding}.

The effectiveness of ECC improves with increasing codeword length $n$.
The design of effective ECC codes
and corresponding decoding processes for long codes is a non-trivial task.
Basing the encoding and decoding operations on storage and retrieval of codewords from
a look-up table quickly becomes impractical as $n$ increases, 
and instead, coding and decoding must be performed algorithmically.
Those algorithms are typically based on complex mathematical theory and
involve substantial computational complexity.
Fortunately it is the decoding complexity that dominates, and that takes place
at the receiver 
where mass and power consumption are not constrained and
processing need not be real-time.

\subsection{Coding for outages}
\label{sec:outageCoding}

Outages are inevitable in optical communication through the atmosphere (see \secref{outages}).
To effectively and efficiently deal with that, 
the \emph{statistics} of the outages must be accurately known at design time.
It is necessarily assumed that
the \emph{actual occurrences} (time and duration)
are stochastic and unknown to the transmitter.

The ECC can be designed to accommodate outages without loss of data reliability.
Under these assumptions,
if the outage probability is $p_O$, then
data can be reliably recovered at an average rate \ile{(1 - p_O) \cdot \rateData{}}
when the ECC is designed with this in mind.
Thus outages that conform to statistical assumptions made at design time do not compromise
reliable data recovery, but do reduce the volume of data that can be reliably recovered.

A useful rule of thumb is that the integrity of data units (such as a single image)
can be recovered reliably
if that data's transmission is spread over a time duration that is at least $50\times$ the longest 
statistically credible outage interval, and this is a constraint on the downlink operation time 
(see \secref{downlinkOperationTime}).
\begin{example}
Consider a set of images that we wish to download.
If they are sent sequentially it would be difficult to recover an image that happened
to coincide with a long outage (such as a week-long cloud cover).
If the longest outage (say from weather events) is expected to be one week, then
each image should be transmitted over a period of 50 weeks (about a year).
Multiple images can be interleaved and transmitted concurrently
to fill out the available $\rateData{}$.
\end{example}

\subsection{ECC efficiency}

The analysis of \S \ref{sec:PPM} provides theoretical bounds on achievable BPP with photon counting
and PPM, without constraint on the choice of ECC scheme.
We saw that specializing the
modulation to PPM results in a small efficiency loss compared to an unconstrained modulation choice.
In a similar manner, the choice of a specific (practical) ECC scheme will introduce a further
efficiency loss.
It will generally also alter the value of \ile{\countPhotonPulse} at which BPP is maximized.
Typically a small increase in \ile{\countPhotonPulse} will be required over the PPM optimum.
The goal of the ECC design process is to minimize this efficiency loss, subject to constraints
on code length and decoding complexity.

For illustration purposes, a concrete example of an ECC design follows.
Coding for outages (see \secref{outageCoding}),
although critically important in this application,
is beyond the scope of this tutorial.

\subsubsection{Concrete example ECC design:  A Reed-Solomon code}

The design of a maximally
efficient ECC scheme is a major design exercise
beyond the scope of this tutorial.  
However, for illustrative purposes 
we can turn to a
well-understood ECC scheme, Reed-Solomon (RS) coding\footnote{We do not provide details
here of the structure of RS codes, nor of their encoding/decoding methods.
},
that has been used extensively in
free-space optical communications systems. It was employed in the NASA/JPL bench
demonstration of \ile{\BPP = 13\text{ bits ph}^{-1}} reported in \cite{RefNumber834}.
Here we present example RS coding results to illustrate how BPP and the
optimal \ile{\countPhotonPulse} are affected by ECC code choice.

RS codes are ``non-binary'' in the sense that the code alphabet is not bits but rather symbols defined
over a finite (Galois) field of size $q$.  Each symbol can be represented uniquely by $m$ bits, where $2^m = q$.
Working in conjunction with $M$-ary PPM where \ile{M = 2^m}, each RS code symbol maps 
naturally onto an individual PPM symbol.
RS codes are particularly effective at overcoming erasures, 
which are the dominant concern
for photon-counting detection in the photon-starved regime.

Consider an RS\ile{\{ n,k,m \}} code, which operates on $m$-bit symbols and encodes $k$ information symbols to produce
codewords of length $n$ symbols.  
The number of redundant symbols is $(n-k)$.  Reed-Solomon codes have the property
that they can recover up to $(n-k)$ erasures occurring across the $n$-symbol codeword.
Thus as we decrease $k$ the redundancy increases and the tolerance of erasures improves.

While the theoretical bounds on $\BPP$ of \secref{PPM} apply to an arbitrary reliability objective,
for a concrete ECC design the calculation of efficiency (in terms of overall $\BPP$) is meaningful only
in the context of a specific decoding error rate requirement.  The required error rate will
decrease as the reliability objective becomes more stringent.

\begin{example}
For example, a reasonable reliability objective might be to achieve an error rate after
decoding that would result in a 1 Mbit image file being received error-free with 99\% likelihood.
\end{example}

In calculating the overall BPP for this objective we must factor in the level of redundancy of the code.
As that redundancy increases, the tolerance to erasures increases (allowing smaller \ile{\countPhotonPulse})
but fewer scientific data bits are conveyed in each codeword.  BPP falls off at high redundancies for this
reason.  It also falls off at low redundancies because higher \ile{\countPhotonPulse} is needed to maintain
the overall error rate objective.  The optimum will fall at some intermediate level of redundancy - although
we know this must be towards the higher end when operating in the photon-starved regime.

\begin{example}
Consider an RS(1023,k,10) code employed with $m=10$ PPM (i.e. 1024-PPM).
In \figref{BPP_vs_code_rate_RS_1023_k_10} we plot the maximum achievable BPP as a function of the code rate $k/n$.
The BPP for each \ile{1 \le k \le 1022} is the value corresponding to the maximum
erasure probability (minimum \ile{\countPhotonPulse}) that still permits the overall reliability
objective to be achieved at that code rate (i.e., the 99\% image reception probability of the previous example).
\end{example}
 
The shortfall in achieved BPP for a concrete EC coding-decoding algorithm relative to the
theoretical least upper bound of \eqnref{PPMcapacity} is called the error-correction \emph{coding efficiency}.
\begin{example}
In the example above we see that the maximum BPP falls short of the 
theoretical limit of \ile{8.62 \text{ bits ph}^{-1}}, as reported in
\tblref{parametersVsBPP2} for $m=10$. 
Thus the coding efficiency is \ile{6.83/8.62 = 0.79\%}.
\end{example}
Efficiencies greater than 0.9 are achievable using more powerful ECC codes.

\incfig
	{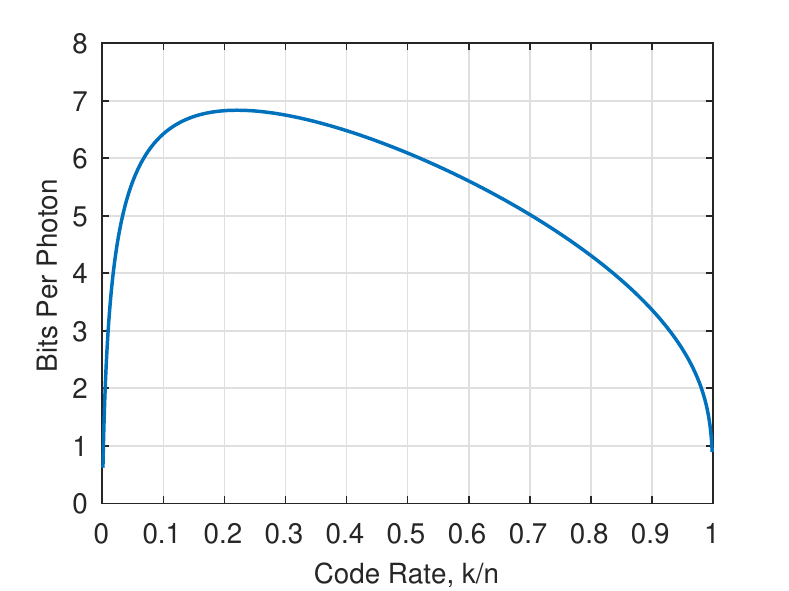}
	{width=.8\linewidth}
	{The maximum achievable BPP as a function of code rate $k/n$ for 
    RS(1023,$k$,10) coded 1024-PPM, assuming random erasures.
	}

The dependence of the optimum \ile{\countPhotonPulse} on the choice of ECC scheme is further illustrated
through \figref{BPP_vs_Ks_RS_1023_k_10} where, for the same RS code as the example, we plot the maximum BPP
as a function of \ile{\countPhotonPulse}.
For each \ile{\countPhotonPulse} the code rate that maximizes BPP has been selected.  For this specific code,
the peak of BPP=6.83 occurs at \ile{\countPhotonPulse=0.32} and a code rate of 0.22.
That these values differ slightly from the theoretical 
least upper bound of \tblref{parametersVsBPP2} is a manifestation of the inefficiency of the selected example RS code
in comparison to an idealized code.

\incfig
	{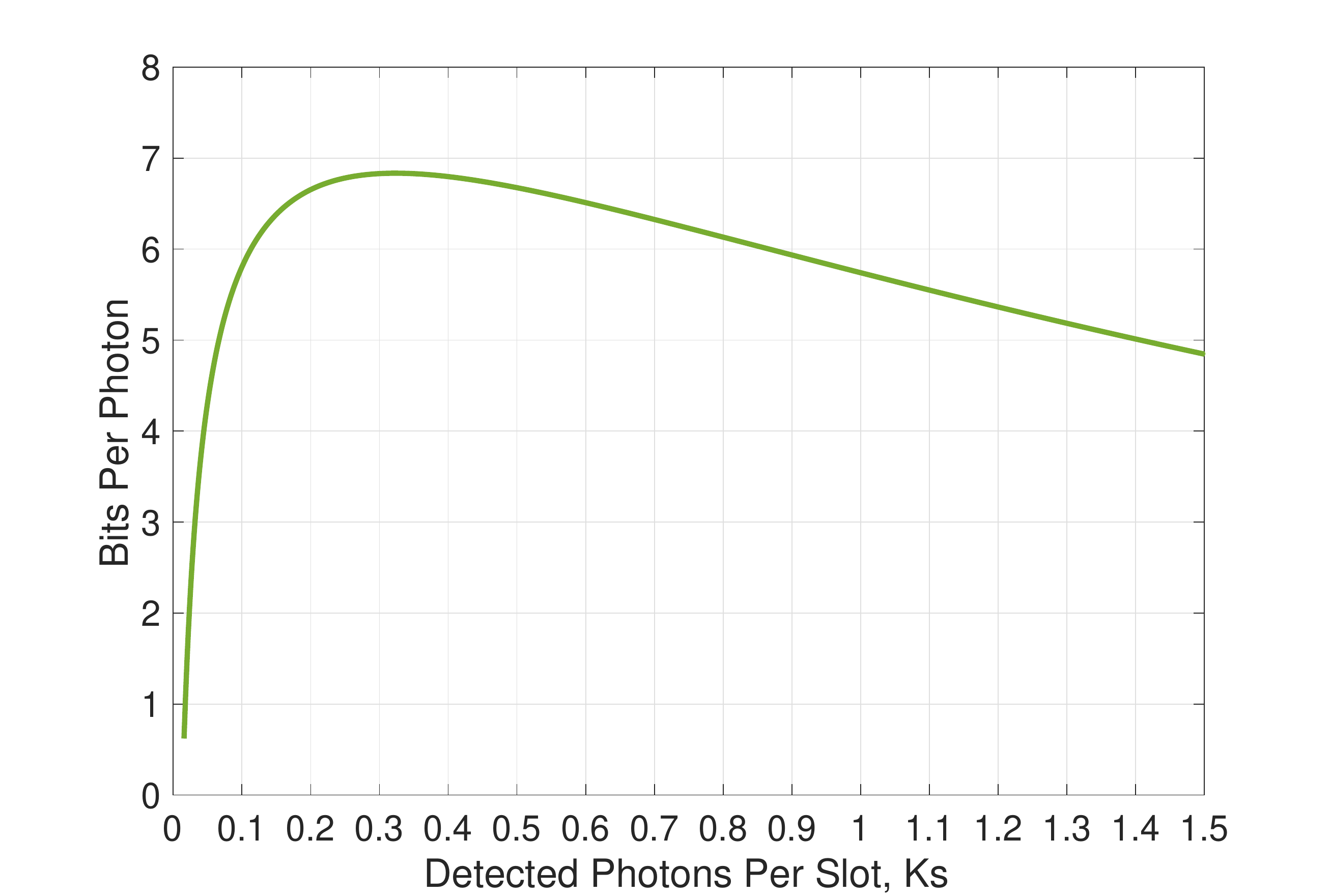}
	{trim=50 0 110 10,width=.75\linewidth}
	{The maximum achievable BPP as a function of \ile{\countPhotonPulse} for 
    RS(1023,$k$,10) coded 1024-PPM, assuming random erasures.
	}

\section{Conclusions}

The end-to-end system design of a communication link, especially one in the context of
interstellar distances and a variety of background radiations of both cosmic and local origin,
is a complex undertaking.
This is due to the significant dependencies among different subsystem and element designs
across transmitter and receiver and TRA and PHY, and between multitudinous performance goals,
theoretical limits, and concrete design choices.
This tutorial has attempted to capture these dependencies as well as the local considerations coming into
play within each subsystem.
A required scope of core principles is large,
spanning quantum mechanics, optics, and device physics on the one hand to
information theory and finite field algebra on the other.
Such an undertaking is best conducted as a collaboration among different types of expertise.
In setting requirements and making concrete tradeoffs, that collaboration should include
the ultimate stakeholders, which includes funding sources and domain scientists.

\section*{Acknowledgements}

The research represented in this tutorial review was supported by a grant from
the Breakthrough Foundation and its Breakthrough StarShot program,
NASA grants NIAC Phase I DEEP-IN – 2015 NNX15AL91G and 
NASA NIAC Phase II DEIS – 2016 NNX16AL32G and the 
NASA California Space Grant NASA NNX10AT93H, 
the Emmett and Gladys W. Technology Fund, and the Limitless Space Institute.
Tom Mozdzen of Arizona State University provided numerous valuable comments and suggestions. 
 
\appendix

\section{Photon efficiency bounds and bandwidth}
\label{sec:optimization}

\subsection{Channel capacity}

The theoretical least upper bound on $\BPP$ is called the \emph{channel capacity}
in information theory
\cite{RefNumber658,RefNumber172,RefNumber171}.
This can be determined in principle for any \emph{channel model}, which mathematically
captures the stochastic relationship between an output waveform or sequence of values
and an input waveform or sequence of values.
As the model embodies more assumptions or restrictions, generally the capacity will decrease.
The channel capacity theorem states that the capacity is the least-upper bound on $\BPP$
over all feasible implementations of transmitter and receiver.

For interstellar communication, three increasingly restrictive models are relevant.
\begin{description}
\item{\textbf{Holevo bound:}}
The most general is a channel in which the transmitter and receiver are allowed to
manipulate multiple quantum states, but without 
interference from background radiation nor quantum entanglement.
This leads to the Holevo bound (see \secref{holevo}).
\item{\textbf{Photon counting:}}
This channel model assumes
conventional electromagnetic transmission with
photon-counting detection in the receiver, and with SBR large enough that background
radiation counts can be neglected.
In this case the channel capacity is
\ile{\BPP \le \log_2 \PAR}, where $\PAR$ is the ratio of peak power to average power \cite{RefNumber799}.
PPM falls in this category with the value
\ile{\PAR = \numberSlots}, and hence its photon efficiency is bounded by 
\ile{\BPP \le \log_2 \numberSlots}.
\item{\textbf{PPM modulation code:}}
The transmitter is constrained to energize exactly one
time slot within each repeated PPM frame.
This reduces the capacity relative to the general photon-counting channel
(see \secref{capacityPPM}).
\end{description}

\newcommand{\numberModes}{N}
\newcommand{\cppa}{\text{PPD}}
\newcommand{\PPD}{\text{PPD}}

\subsection{Quantum limit}
\label{sec:holevo}

For purposes of computing the quantum limit, one channel use
is allocated a fixed number of photon detections
partitioned across $\numberModes$ modes or dimensions,
each mode occupying a non-overlapping time-wavelength space.
Define \emph{photons per dimension} \ile{\PPD} as the
average number of detected photons per dimension.
The resulting capacity-achieving quantum efficiency is the so-called Holevo capacity \cite{RefNumber800}
\begin{equation}
\label{eq:HolevoCapacity}
\BPP = (1 + 1/\PPD) \cdot \log_2 (1+\PPD) - \log_2 \PPD
\,.
\end{equation}
This is based on the statistics predicted by non-entangled quantum states, where it is assumed that
the transmitter and receiver can manipulate and read their local quantum state directly.
It is also assumed, for the purposes of channel coding, that an unlimited number of
quantum states can be manipulated and read jointly and collectively.
$\BPP$ vs $\PPD$ is plotted as the quantum limit in \figref{opticalLimitsQuantum}.
While there is no available technology to approach this limit, it represents a theoretical ideal
that quantifies the opportunities for more advanced future technologies.

\incfig
	{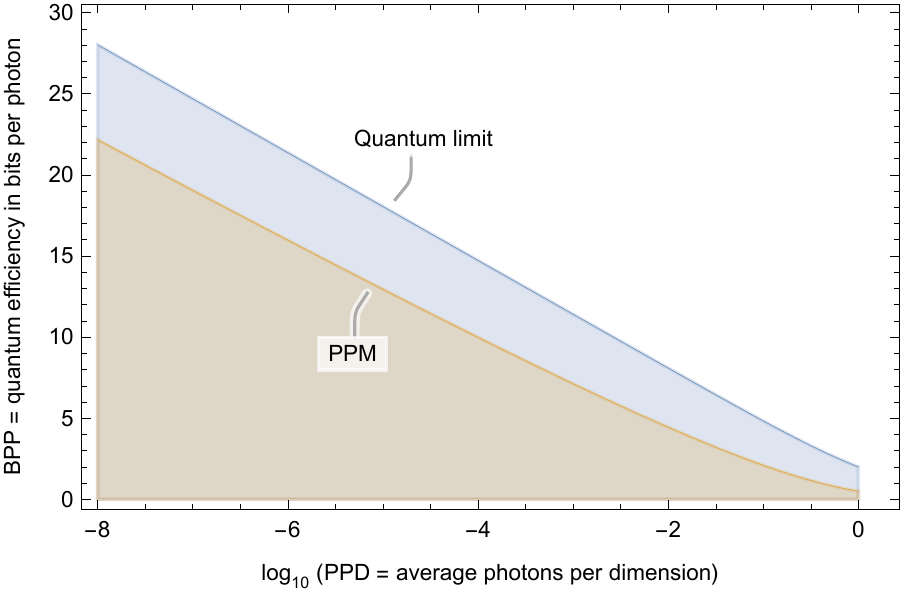}
	{
	trim=0 0 0 0,
    	clip,
    	width=.7\linewidth
	}
	{
The least upper bound on photon efficiency $\BPP$ of \eqnref{HolevoCapacity} is plotted against
	 the logarithm of average detected photons per dimension $\PPD$.
	 Any point in the shaded region is theoretically achievable, although the most advantageous
	 choice is on the upper boundary.
	 Achieving larger $\BPP$ is associated with a smaller $\PPD$ (which is called photon starvation), 
	 which requires spreading the total photon count
	over an increasing number of quantum modes or dimensions.
	This is compared against PPM as given by \eqnref{capacityPPM},
	quantifying the `headroom' available for future technology advances that eliminate
	the PPM photon-counting limitations and manipulate the quantum states directly.
}

If we have a fixed average total number of photons across all dimensions, then we can force
\ile{\PPD \to 0} as \ile{\numberModes \to \infty}, and in that case \ile{\BPP \to \infty} in \eqnref{HolevoCapacity}.
The primary lesson to draw from \figref{opticalLimitsQuantum} is 
that high $\BPP$ is associated with \emph{photon starvation},
in which $\PPD$ is very small.
Both the reasoning behind this, as well as the consequences for data link design, are subtle.
First the logic: As the detected photons per channel use decreases, 
the data bits that can be recovered per channel use also decreases.
This would seem to be a problem, except
that the transmitted and received energy per channel use is reduced by an even larger
factor, which is why $\BPP$ improves.
The fewer data bits recovered per channel use can be offset by employing more channel uses,
which, as we will see shortly, expands the optical bandwidth.

\newcommand{\timeChannelUse}{\mathcal T}
\newcommand{\numberTimeSlots}{L}
\newcommand{\numberFrequencies}{J}

\subsubsection{Quantum efficiency vs bandwidth expansion}
\label{sec:bwExpansionQuantum}

Not surprisingly photon starvation is intricately related to bandwidth expansion,
since it requires a growth in time-frequency $\numberModes$ for each channel use.
Increasing the number of times expands bandwidth because this requires shorter mode time
durations.
We now establish that the minimum bandwidth is
\begin{equation}
\label{eq:BWtoRho}
\frac{\bandwidth}{\rateData{}} = \frac{1}{\PPD \cdot \BPP} \,,
\end{equation}
leading directly from \figref{opticalLimitsQuantum} to \figref{opticalLimitsRho}.
This relation is difficult to interpret directly because $\PPD$ and $\BPP$
have a complicated relationship captured for example in \eqnref{HolevoCapacity}.

The derivation of \eqnref{BWtoRho} follows by considering the smallest feasible bandwidth.
Let the time associated with one channel use be fixed at $\timeChannelUse$.
A set of non-overlapping time-wavelength slots constitute the modes or dimensions.
Let there be \ile{\numberModes = \numberTimeSlots \cdot \numberFrequencies}
such modes, where $\numberTimeSlots$ is the number of unique times and
$\numberFrequencies$ is the number of frequencies.
Then the time duration of each mode is at most \ile{\timeChannelUse/\numberTimeSlots},
the bandwidth of each mode is at least \ile{\numberTimeSlots/\timeChannelUse},
and the total bandwidth is
\ile{\bandwidth = \numberFrequencies \cdot \numberTimeSlots/\timeChannelUse = \numberModes/\timeChannelUse}.
The number of bits recovered per channel use is $\rateData{} \timeChannelUse$
and the average number of detected photons per channel use is $\numberModes \cdot \PPD$.
Hence  \eqnref{BWtoRho} is confirmed since
\begin{equation}
\BPP = \frac{\rateData{} \timeChannelUse}{\numberModes \cdot \PPD}
=  \frac{\rateData{}}{\bandwidth \cdot \PPD}
\,.
\end{equation}

\subsection{PPM modulation code}
\label{sec:capacityPPM}

Specializing to a discrete-time channel model constrained to
the statistically independent repetition of  PPM frames,
the capacity is given by \eqnref{PPMcapacity}.
The PPM parameters \ile{\{\countPhotonPulse,\numberSlots\}} 
can be chosen optimally to maximize \eqnref{PPMcapacity} as follows.
In this case \ile{\cppa = \countPhotonPulse/\numberSlots},
which equals the number of photon detections averaged over all $\numberSlots$ time slots.
Eliminating $\countPhotonPulse$ in \eqnref{PPMcapacity} and
maximizing the bound with the optimal choice of $\numberSlots$,
\begin{equation}
\label{eq:capacityPPM}
\BPP \le \max_{\numberSlots \ge 2} 
\bigg\{ \alpha \big[ \numberSlots \cdot \PPD \, \big] \cdot \log_2 \numberSlots \bigg\}
\,.
\end{equation}
This equation yields a relationship between $\BPP$ and \ile{\{\countPhotonPulse,\numberSlots\}} parameterized by
$\PPD$.
This optimization is illustrated graphically in Fig.27 of \cite{RefNumber833}.
Computing this relationship for different values of $\PPD$ yields the PPM boundaries in
 \figref{opticalLimitsRho}, \figref{photonStarvation}, and \figref{opticalLimitsQuantum}. 
\end{makefigurelist}

\end{document}